\documentclass[12pt,preprint]{aastex}

\newcommand{\eight}{850~$\mu$m}
\newcommand{\four}{450~$\mu$m}

\newcommand{\ks}{Kolmogorov-Smirnov}
\newcommand{\boldrho}{\mbox{\boldmath$\rho$}}

\begin{document}

\title{High Mass Star Formation. III. The Functional Form of the 
Submillimeter Clump Mass Function}

\author{Michael A. Reid and Christine D. Wilson} \affil{Department of 
Physics and Astronomy, McMaster University, Hamilton, ON, L8S 4M1, Canada 
\& Harvard-Smithsonian Submillimeter Array Project, 645 N A'ohoku Pl., 
Hilo, 
HI, 96720, USA}

\begin{abstract} We investigate the mass function of cold, dusty clumps in 
11 low- and high-mass star-forming regions.  Using a homogeneous fitting 
technique, we analyze the shape of each region's clump mass function and 
examine the commonalities among them.  We find that the submillimeter 
continuum clump mass function in low-mass star-forming regions is 
typically best fit by a lognormal distribution, 
while that in high-mass star-forming regions is better 
fit by a double power law.  A single power law clump mass distribution is 
ruled out in all cases.  Fitting all of the regions with a double 
power law, we find the mean power law exponent at the high-mass end of 
each mass function is $\alpha_{\rm high} = -2.4\pm0.1$, consistent with 
the Salpeter result of $\alpha = -2.35$.  We find no region-to-region 
trend in $\alpha_{\rm high}$ with the mass scale of the clumps in a given 
region, as characterized by their median mass.  Similarly, non 
non-parametric tests show that the \emph{shape} of the clump mass function 
does not change much from region to region, despite the obvious changes in 
the intrinsic mass scale.  This result is consistent with the hypothesis 
that the clump mass distribution is determined by a highly stochastic 
process, such as turbulent fragmentation.  It may also suggest that the 
data reduction and analysis techniques strongly affect the shape of the 
derived mass function. 
\end{abstract}


\section{INTRODUCTION}

	Advances in submillimeter continuum imaging techniques in the past 
decade have dramatically improved our understanding of the structure of 
the dense, cold interstellar medium (ISM) in star-forming regions.  In 
every region studied, the cold, dusty ISM exhibits a clumpy, often 
filamentary structure whose density maxima correspond to sites of present 
or future star formation activity.  The primary metric used to describe 
this inhomogeneous structure is the clump/core\footnote{Throughout this 
paper, we reserve the term ``core'' to denote the small 
(diameter~$\lesssim 0.1$~pc), dense condensations thought to be the 
immediate precursors of individual stars or small multiple systems.  We 
use the term ``clump'' more generally to denote discrete structures which 
may form zero, one, or many stars and may therefore be much more massive 
than ''cores''.  In this sense, cores may be considered a subset of 
clumps.} mass function.  Submillimeter continuum observations of the core 
mass function in nearby regions of low-mass star formation, such as 
Serpens \citep{ts98}, $\rho$~Oph \citep{man98,dj2000b}, and Orion~B 
\citep{m01,dj2001}, have provided compelling evidence that the mass 
function of cold, dense cores mirrors the stellar initial mass function 
(IMF, e.g. \citealt{kroupa02}, \citealt{chabrier03}).  It appears, 
therefore, that these dusty cores are the immediate precursors of stars.  
Two main arguments are typically advanced in favor of this interpretation.  
First, both the stellar and core mass functions are well fit by 
Salpeter-like power laws \citep{sal55} above $\sim$1~$M_{\odot}$.  
Second, both the core and stellar mass functions appear to flatten out 
below $\sim$1~$M_{\odot}$ and peak at about 0.1~$M_{\odot}$ 
\citep{ts98,man98,dj2000b,dj2001,m01,kroupa02}.  In other words, the core 
mass function and the stellar IMF have similar \emph{shapes} and intrinsic 
mass \emph{scales}.  Thus, by applying a more or less constant 
core-to-star mass conversion efficiency, the core mass function can be 
converted into the stellar IMF.

	In papers~I and II in this series, we showed that, despite the 
large difference in the median clump mass, the clump mass function in 
massive star-forming regions NGC~7538 and M17 has a similar shape to that 
in low-mass star-forming regions (\citealt{paperi,paperii}; Papers~I and 
II hereafter).  In both NGC~7538 and M17, we found that double power law 
fits to the clump mass functions produce exponents which are consistent 
with the Salpeter mass function \citep{sal55}.  The similarities in their 
shapes suggests that the clump mass functions in low- and high-mass 
star-forming regions may have similar origins.  What if the mass function 
of massive clumps were found to resemble that of massive stars, as is the 
case for low-mass stars?  That would suggest that massive clumps could 
also be converted to massive stars (or small multiple systems) on a more 
or less one-to-one basis.  Such a finding would lend weight to theories 
which suggest that massive stars form by the collapse of individual clumps 
\citep{mt02,mt03,krum05,krum06}, rather than those which suggest that 
massive stars form by coalescence or competitive accretion 
\citep{bon97,bon01,bon04,bon06}.

	To address these important matters, it is necessary to be able to 
accurately fit and interpret clump mass functions.  In Paper~I, we argued 
that variations in analytic technique might account for much of the 
observed variation in the power-law exponents derived from fits to 
submillimeter continuum clump mass functions.  In particular, we suggested 
that the systematic differences between the power-law exponents of clump 
mass functions in low- and high-mass regions probably result from the fact 
that the former are typically fit with two power law segments, while the 
latter are usually fit with only one (compare \citet{m01} and 
\citet{dj2000b} with \citet{tot02} and \citet{mookerjea04}).  In this 
paper, we extend and develop these arguments using a more thorough study 
of the available data on the submillimeter continuum clump mass function.  
In \S\ref{sec:howtofit}, we discuss how best to fit a clump/core mass 
function, paying special attention to means by which the fits may be 
misinterpreted.  In \S\ref{sec:funcform}, we apply a consistent fitting 
methodology to determine which of several plausible functions best 
represents the clump/core mass function in a variety of star-forming 
regions.  Finally, in \S\ref{sec:compare}, we use non-parametric tests to 
determine whether the shape of the clump/core mass function varies 
significantly from region to region or changes with the mass scale 
studied.

\section{FITTING AND INTERPRETING CLUMP MASS FUNCTIONS}
\label{sec:howtofit}

	Clump mass functions are commonly expressed in two forms:  
differential and cumulative.  The differential mass function (DMF), 
$\Delta N/\Delta M$, has the advantages of permitting a simple Poisson 
uncertainty analysis but suffers the arbitrariness of binning.  Binning 
can be a serious impediment to the accurate interpretation of stellar mass 
functions \citep{scalo98,apellaniz}.  Thus, the differential mass function 
is not suitable for use when the total number of clumps, $N_{\rm cl}$, is 
small.  The cumulative mass function (CMF), $N(>M)$, requires a more 
complicated uncertainty analysis, but it suffers none of the problems 
associated with binning and therefore is suitable even in the small 
$N_{\rm cl}$ regime.  The CMF is therefore preferable for studies of the 
submillimeter clump/core mass function where the total number of objects 
is typically $\lesssim 100$.  However, the CMF is easily misinterpreted, 
as we now demonstrate.

	Consider a very simple mass function whose DMF is well-fit by a 
single power law of the form

\begin{equation}
\frac{\Delta N}{\Delta M} = AM^{\alpha}~~, \label{eq:dmfeq}
\end{equation}

\noindent where $\Delta N$ is the number of objects in a mass bin of width
$\Delta M$, and $A$ and $\alpha$ are constants.  It is sometimes assumed
that the complementary cumulative mass function is also a power law whose
exponent differs from $\alpha$ by 1:

\begin{eqnarray}
N(>M) &=& \int_{M}^{\infty} \frac{\Delta N}{\Delta M} dM \nonumber \\  
      &=& -\frac{A}{\alpha+1}M^{\alpha+1}~~, \label{eq:pl1}
\end{eqnarray}

\noindent for $\alpha < -1$.  The expectation is therefore that, if the 
DMF is well represented by a single power law, the CMF will be too.  In 
measured mass functions, this may not be the case.  A measured mass 
function contains a finite sample of clumps with an upper mass limit, 
$M_{\rm max}$.  This upper mass limit can be either a real cutoff on the 
mass distribution or the result of finite sampling.  In either case, the 
DMF will remain the simple power law of Equation~\ref{eq:dmfeq}, but the 
CMF will deviate from Equation~\ref{eq:pl1}.  Integrating the DMF of 
Equation~\ref{eq:pl1} to an upper limit of $M_{\rm max}$, we obtain

\begin{eqnarray}
N(>M) &=& \int_{M}^{M_{\rm max}} \frac{\Delta N}{\Delta M} dM \nonumber \\
      &=& -\frac{A}{\alpha+1}M^{\alpha+1} + \frac{A}{\alpha+1}M_{\rm max}^{\alpha+1}~~.\label{eq:pl2}
\end{eqnarray}

\begin{figure}
\begin{center}
\includegraphics[width=6.0in]{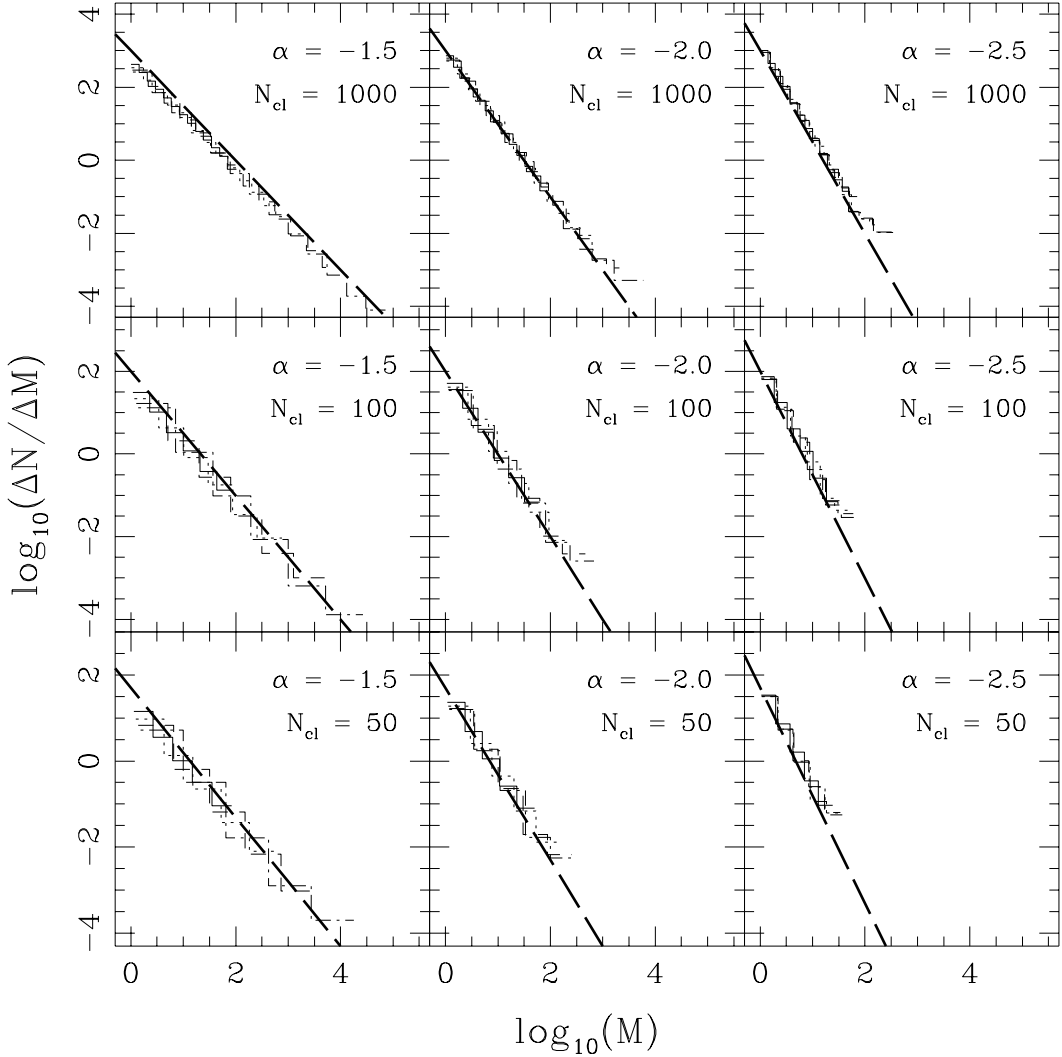}
\caption{Differential mass functions from simulated clump mass
distributions.  Each simulated data set consists of N$_{\rm cl}$ clumps
drawn randomly from a mass function of the form $\Delta N/\Delta M
\propto M^{\alpha}$ (\emph{thick dashed line}), with N$_{\rm cl}$ and
$\alpha$ values as shown.  To produce each plotted DMF, 1000 simulated
data sets were averaged together.  The four simulated DMFs plotted in
each panel vary in the range of masses from which the simulated masses
were drawn: 1--100 (\emph{solid line}), 1--1000 (\emph{dotted line}),
1--10$^{4}$ (\emph{short dashed line}), and 1--10$^{5}$ (\emph{dot-dashed
line}).  The mass units are arbitrary.  The number of bins in each mass
function is set to $(2{\rm N}_{\rm cl})^{1/3}$. 
\label{fig:fake_dmfs}}
\end{center} 
\end{figure}

\begin{figure}
\begin{center}
\includegraphics[width=6.0in]{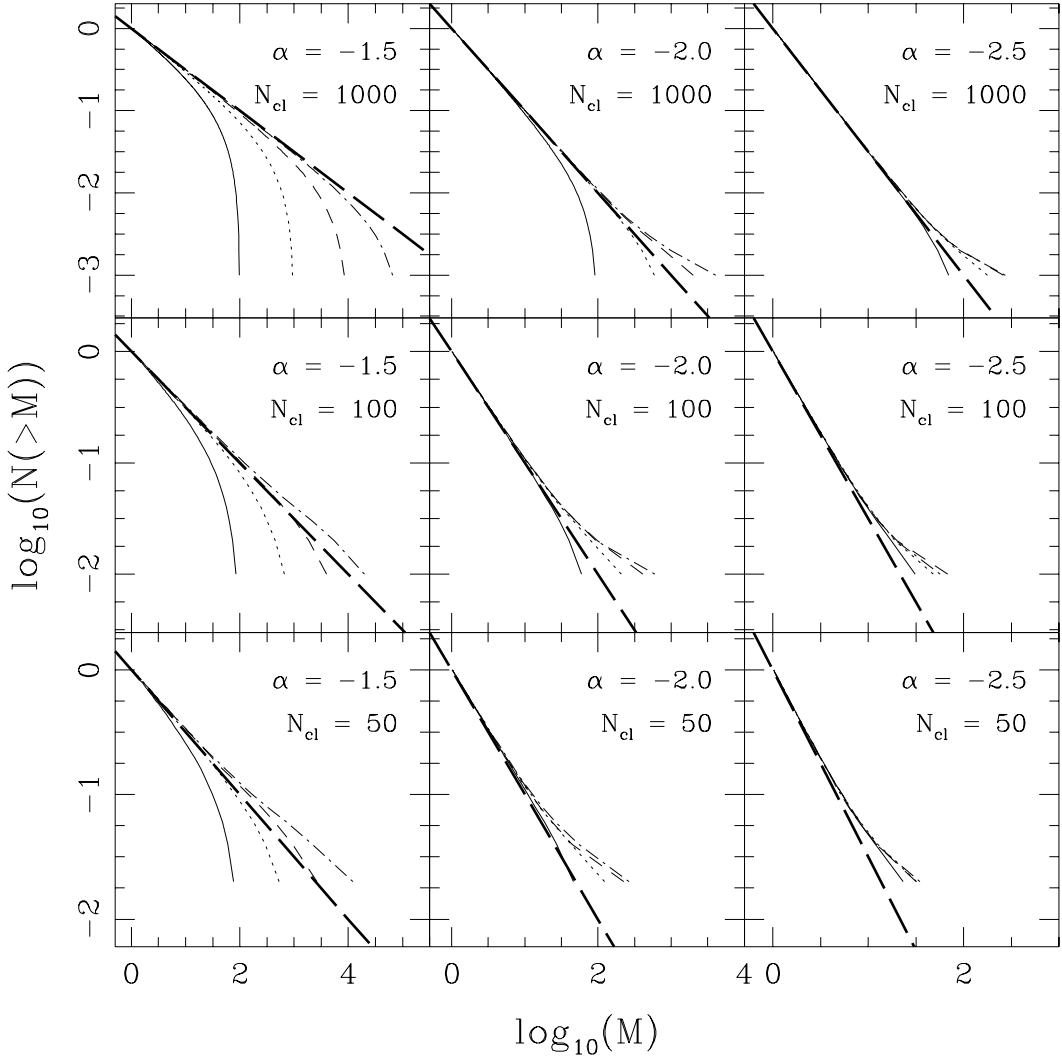}
\caption{Cumulative mass functions from simulated clump mass
distributions, corresponding to those whose differential mass functions
are shown in Fig.~\ref{fig:fake_dmfs}.  As in Fig.~\ref{fig:fake_dmfs},
the line types indicate the clump mass range from which simulated masses
(arbitrary units) were drawn:  1--100 (\emph{solid line}), 1--1000 
(\emph{dotted line}), 1--10$^{4}$
(\emph{short dashed line}), and 1--10$^{5}$ (\emph{dot-dashed line}).  The
thick dashed line represents the power law distribution from which the
clump masses were drawn.  For sufficiently steep power laws ($\alpha =
-2.5$), the CMF is likely to be well-fit by a single power law, even when
the total number of clumps is small.  However, for shallower power laws
($\alpha = -1.5$) and CMFs spanning smaller clump mass ranges, the plotted
CMF deviates significantly from the single power law form.
\label{fig:fake_cmfs}}
\end{center}
\end{figure}

The practical significance of the difference between 
Equations.~\ref{eq:pl1} and \ref{eq:pl2} depends on the range of clump 
masses (i.e. the value of $M_{\rm max}/M_{\rm min}$), the value of the 
power-law 
exponent, $\alpha$, and the number of clumps in the sample, N$_{\rm cl}$.
In Figures~\ref{fig:fake_dmfs} and \ref{fig:fake_cmfs}, we use simulated 
data sets to show how the shape of both the DMF and CMF depend on each of 
these parameters.  To simulate a clump mass function, we randomly draw 
N$_{\rm cl}$ masses from the interval $M_{\rm min}$ to $M_{\rm max}$ 
following the distribution $\Delta N/\Delta M \propto M^{\alpha}$.  We do 
this for representative values of $N_{\rm cl}$ = 50, 100, and 1000.  
We simulate power law exponents of $\alpha$ = -1.5, 
-2.0, and -2.5.  This range encompasses the exponents of the Salpeter IMF 
($\alpha$ = -2.35; \citealt{sal55}), the CO clump mass function ($\alpha = 
-1.7$; \citealt{kramer98}), and the values typically found in 
submillimeter continuum clump surveys.  We vary the ratio $M_{\rm 
max}/M_{\rm min}$ by holding $M_{\rm min}$ constant at 1 (the mass units 
are arbitrary) and taking $M_{\rm max}$ values of $10^{2},10^{3},10^{4}$, 
and 10$^{5}$.  For each combination of N$_{\rm cl}$, $\alpha$, and $M_{\rm 
max}$, we produce a representative data set by averaging 1000 
randomly-generated sets.  Figures~\ref{fig:fake_dmfs} and 
\ref{fig:fake_cmfs} show the resulting DMFs and CMFs, respectively.  Note 
that we follow the popular convention of normalizing $N(>M)$ by N$_{\rm 
cl}$.

	By construction, the DMFs shown in Figure~\ref{fig:fake_dmfs} all 
follow the input mass distributions.  However, as shown in 
Figure~\ref{fig:fake_cmfs}, the CMFs clearly deviate from the power law 
form suggested by Equation~\ref{eq:pl1}.  As seen in the right column of 
Figure~\ref{fig:fake_cmfs}, a steep power law ($\alpha=-2.5$) overwhelms 
the effects of both $N_{\rm cl}$ and $M_{\rm max}$, recovering essentially 
the pure single power law in all cases.  As seen in the middle and left 
columns of Figure~\ref{fig:fake_cmfs}, however, shallower power laws 
($\alpha=-1.5,-2.0$) often cause the CMF to deviate from the pure power 
law behaviour expected from Equation~\ref{eq:pl1}.  This is especially the 
case when the ratio of $M_{\rm max}/M_{\rm min}$ is small (solid line in 
Fig.~\ref{fig:fake_cmfs}).  It is therefore likely that fitting real data 
sets with single power laws of the form of Equation~\ref{eq:pl1} would 
give highly misleading power law exponents.  To achieve accurate results, 
fitted functions must account for the maximum clump mass, as in 
Equation~\ref{eq:pl2}.

	Many of the simulated CMFs shown in Figure~\ref{fig:fake_cmfs} 
strongly resemble the CMFs derived from observations of real star-forming 
regions.  To account for their curvature, these observational CMFs are 
typically fitted with two power law segments (e.g. \citealt{dj2000b}).  
That intrinsically single power-law CMFs should so closely resemble those 
typically fit with two power laws implies that great care should be taken 
in interpreting the shape of the CMF and in choosing functions with which 
to fit it.  In simple cases, such as the single power-law simulated here, 
a check of the DMF should help constrain the functional forms which might 
be fitted to the CMF.  Even in this simple case, which is simpler than is 
likely to be encountered in real data, care must be taken when fitting the 
CMF in order to recover the correct power-law exponent: accounting for 
$M_{\rm max}$ is clearly essential.  In CMFs measured from real 
observations, the effects of incompleteness, small-number statistics, and 
binning can all lead to uncertainty about the functional form of the clump 
mass function.  We believe that particular care is warranted where the 
observed clump/core mass function encompasses the peak in the stellar IMF 
at $\sim$0.1~$M_{\odot}$ \citep{kroupa02,chabrier03}.  In such cases, 
incompleteness and a real physical turnover in the mass function are 
likely to be difficult to distinguish.

\section{THE FUNCTIONAL FORM OF THE CLUMP MASS FUNCTION}
\label{sec:funcform}

	We now turn to the question of the functional form of the 
clump/core mass function.  The form most commonly employed in studies of 
low-mass star-forming regions is a double power law above 
$\sim$0.1$M_{\odot}$ with a break between the power law segments at 
$\sim$0.5$M_{\odot}$ \citep{man98,dj2000b,m01,dj2001}.  This is 
essentially the same form as found for the Galactic field single-star IMF 
\citep{kroupa02}.  However, a double power law is not the only plausible 
shape for the clump mass function.  In Paper~II, we showed that the M17 
clump mass function is better fit by a lognormal distribution than by a 
double power law.  Moreover, in \S\ref{sec:howtofit} above, we noted that 
the finite clump mass range in any real clump mass function adds curvature 
to the CMF.  This curvature can make a mass spectrum which is 
intrinsically a single power law appear to adopt a shape similar to those 
usually fit with a double power law.  Additional uncertainties about the 
shape of the clump/core mass function arise due to the diversity of 
techniques used by different authors to fit such mass functions.

	To eliminate the effects of differing fitting techniques, we have 
gathered all of the published lists of clump/core masses from unbiased 
millimeter and submillimeter continuum surveys of star-forming regions and 
fit them in a homogeneous manner.  In Table~\ref{tab:studies}, we 
summarize these 11 different clump/core mass functions measured in 7 
different star-forming regions.  In cases where the original authors used 
more than one set of assumptions about the dust temperature when 
calculating the clump masses, we use the masses calculated assuming a 
uniform dust temperature, which is the method most widely used.  The only 
exceptions are the studies by \citet{man98} and \citet{m01}, which divide 
the clumps into several groups, each with a single dust temperature, 
before calculating their masses.  Not all of the studies made comparable 
attempts to distinguish between starred and starless cores/clumps, so we 
have used the complete list of all objects in each case (except our own 
study of M17 from Paper~II, where we have two comparable data sets in 
which clumps coincident with MSX sources were removed in a consistent 
way).  The clumps in these 7 star-forming regions span more than 5
orders of magnitude in mass (0.05--16000~$M_{\odot}$) and more than two 
orders of magnitude in linear size.  The median clump masses range from 
0.17~$M_{\odot}$ in $\rho$~Oph to 470~$M_{\odot}$ in RCW~106.  For 
comparison with the theory of turbulent fragmentation, we also include in 
our analysis a list of clumps extracted from a computational hydrodynamic 
simulation of 4.6 Jeans masses of turbulent, self-gravitating gas (run B5 
of \citealt{tilley04}, TP04 hereafter).  There are two caveats to any 
comparison made between the TP04 and observational mass functions.  
First, the dimensionless nature of the TP04 simulations means the absolute 
mass scale is arbitrary.  Second, the number of Jeans masses simulated by 
TP04 is considerably less than is found in real star-forming regions, 
which could have consequences for the number and mass distribution of 
fragments formed.

\begin{deluxetable}{clccccccc}
\tabletypesize{\footnotesize}
\tablewidth{0pt}
\tablecaption{Submillimeter and Millimeter Continuum Clump Surveys\label{tab:studies}}
\tablehead{
\colhead{Data Set ID} &
\colhead{Region Name} & 
\colhead{$\lambda$ [$\mu m$]} & 
\colhead{N$_{\rm clumps}$} & 
\colhead{HPBW [pc]} &
\colhead{M$^{\rm median}_{\rm clump}$} & 
\colhead{M$^{\rm max}_{\rm clump}$} & 
\colhead{M$^{\rm tot}_{\rm clump}$} & 
\colhead{Ref.} 
}
\startdata
1 & $\rho$ Oph & 1300 & 62 & 0.0085 & 0.17 & 3.2 & 24 & 1 \\
2 & $\rho$ Oph & 850 & 55 & 0.011 & 0.19 & 6.33 & 29 & 2 \\
3 & Orion B & 850 & 82 & 0.02 & 0.8 & 9.20 & 106 & 3 \\
4 & Orion B & 850 & 75 & 0.03 & 0.86 & 30.34 & 120 & 4 \\ 
5 & Lagoon/M8 & 450 \& 850 & 37 & 0.13 & 6.93 & 34.1 & 287 & 5 \\
6 & M17 & 450 & 96 & 0.07 & 13 & 160 & 1700 & 6 \\
7 & M17 & 850 & 105 & 0.12 & 11 & 120 & 2100 & 6 \\
8 & NGC~7538 & 450 & 77 & 0.11 & 24 & 2700 & 8800 & 7 \\
9 & NGC~7538 & 850 & 67 & 0.21 & 34 & 3000 & 9300 & 7 \\ 
10 & W43 & 1300 & 48 & 0.29 & 135 & 3600 & $1.4\times10^{4}$ & 8 \\
11 & RCW 106 & 1200 & 95 & 0.42 & 470 & $1.6\times10^{4}$ & $9.9\times10^{4}$ & 9 \\
12 & simulation & \nodata & 381 & 0.0014\tablenotemark{a} & 0.045 & 25 & 57\tablenotemark{b} & 10 \\
\enddata
\tablenotetext{a}{Size of a cubic simulation pixel.}
\tablenotetext{b}{Total mass of gravitationally bound clumps found in the simulation.}
\tablerefs{(1) \citet{man98}; (2) \citet{dj2000b}; (3) \citet{m01}; (4) 
\citet{dj2001}; (5) \citet{tot02}; (6) \citet{paperii}; (7) \citet{paperi}; (8) 
\citet{msl03}; (9) \citet{mookerjea04}; (10) \citet{tilley04} }
\end{deluxetable}

\subsection{Functional Forms Fitted}

	In the literature, differential clump mass functions are usually 
fit by either a single power law,

\begin{equation}
\frac{\Delta N}{\Delta M} = AM^{\alpha}
\label{eq:dmf1pl}
\end{equation}

\noindent or a double power law,

\begin{equation}
\frac{\Delta N}{\Delta M} = \left\{ \begin{array}{rl}
 AM_{\rm break}^{(\alpha_{\rm high} - \alpha_{\rm low})}M^{\alpha_{\rm low}} & ,~M < M_{\rm break} \\
 AM^{\alpha_{\rm high}} & ,~M \geq M_{\rm break}
 \end{array} \right.
\label{eq:dmf2pl}
\end{equation}

\noindent where $\alpha_{\rm low}$ and $\alpha_{\rm high}$ are, 
respectively, the power-law exponents below and above the break mass, 
$M_{\rm break}$.  The curvature seen in observed clump CMFs makes it is clear that 
the single power law of Equation~\ref{eq:dmf1pl} cannot be applied to the 
CMF.  However, many authors have made simple double power-law fits to the 
CMF, as per

\begin{equation}
N(>M) = \left\{ \begin{array}{rl}
 AM_{\rm break}^{(\alpha_{\rm high} - \alpha_{\rm low})}M^{\alpha_{\rm low}+1} & ,~M < M_{\rm break} \\
 AM^{\alpha_{\rm high}+1} & ,~M \geq M_{\rm break}~~.
 \end{array} \right.
\label{eq:cmf2pl}
\end{equation}

\noindent Equation~\ref{eq:cmf2pl} is the first function with which we 
will fit observed CMFs, and we call it 2PL.  

	We can derive two other potential fitting functions by integrating 
Equations~\ref{eq:dmf1pl} and \ref{eq:dmf2pl} to obtain their 
corresponding CMFs.  As per the discussion of \S\ref{sec:howtofit}, we 
integrate to a finite upper mass limit, $M_{\rm max}$.  The single power 
law CMF with an upper mass limit, which we call 1PLMM, is given by:

\begin{equation}
N(>M) = \int_{M}^{M_{\rm max}} \frac{\Delta N}{\Delta M}dM = \left\{\begin{array}{rl}
\frac{A}{(\alpha+1)}(M_{\rm max}^{\alpha+1} - M^{\alpha +1}) & ,~M < M_{\rm max}\\
0 & ,~M \geq M_{\rm max}~~.
\end{array} \right.
\label{eq:cmf1plmm}
\end{equation}

\noindent Similarly the double power law CMF with an upper mass limit, 
which we call 2PLMM, is given by:

\begin{equation}
N(>M) = \left\{ \begin{array}{rl}
 \frac{AM_{\rm break}^{(\alpha_{\rm high} - \alpha_{\rm low})}}{(\alpha_{\rm low} +1)}(M_{\rm break}^{\alpha_{\rm low}+1} -M^{\alpha_{\rm low} +1}) \\ + \frac{A}{(\alpha_{\rm high}+1)}(M_{\rm max}^{\alpha_{\rm high}+1} - M_{\rm break}^{\alpha_{\rm high} +1})& ,~M < M_{\rm break} \\
 \frac{A}{(\alpha_{\rm high}+1)}(M_{\rm max}^{\alpha_{\rm high}+1} -M^{\alpha_{\rm high}+1}) & ,~M_{\rm break} \leq M < M_{\rm max} \\
 0 & ,~M \geq M_{\rm max}~~.
 \end{array} \right.
\label{eq:cmf2plmm}
\end{equation}

	Inspired by similar work on the stellar mass function, we also 
consider a lognormal DMF:

\begin{equation}
\frac{\Delta N}{\Delta M} = \frac{1}{A_{1}
\sqrt{2\pi}M}~\mbox{exp}\left[-\frac{(\mbox{ln}M -
A_{0})^{2}}{2A_{1}^{2}}\right]~~,
\label{eq:dmflogn}
\end{equation}

\noindent whose complementary CMF, which we call LOGN, is given by:

\begin{equation}
N(>M) = \frac{1}{2}\left[1 - \mbox{erf}\left(\frac{\mbox{ln}M -
A_{0}}{\sqrt{2}A_{1}}\right)\right]~~.
\label{eq:cmflogn}
\end{equation}

	We constructed a CMF for each region using the published masses.  
To assign weights to each point for use in the calculation and 
minimization of $\chi^{2}$, we make the same assumptions as in Paper~II, 
namely that the mass uncertainties, $\sigma_{M}$, are distributed roughly 
as $\sigma_{M} \propto M$, and that the CMF can be approximately 
represented by a power law.  Under these assumptions, the uncertainties in 
$N(>M)$, $\sigma_{N}$, are approximately given by $\sigma_{N} \propto N$ 
(see Paper~II).  This method of assigning weights avoids the use of 
individual mass uncertainties, which are frequently unavailable.  To 
assess the uncertainty in each fitted parameter, we must again choose a 
method which does not rely on the mass uncertainties.  For each CMF, we 
construct and fit 10$^{5}$ Monte Carlo realizations of the data using the 
bootstrap method (random sampling with replacement, e.g. \citealt{press}).  
Using the resulting distributions of fitted parameters, which are 
typically symmetric about the best-fit values, we construct 95\% ($\sim 
2\sigma$) confidence limits on each parameter and report these as our 
uncertainties.  Using the distribution of $\chi^{2}$ values, we compute 
the probability, $P$, of obtaining by chance a $\chi^{2}$ value poorer 
than that from the best fit to the original CMF.  Higher $P$ values 
indicate a better quality of fit.

\begin{figure}
\begin{center}
\includegraphics[width=\columnwidth]{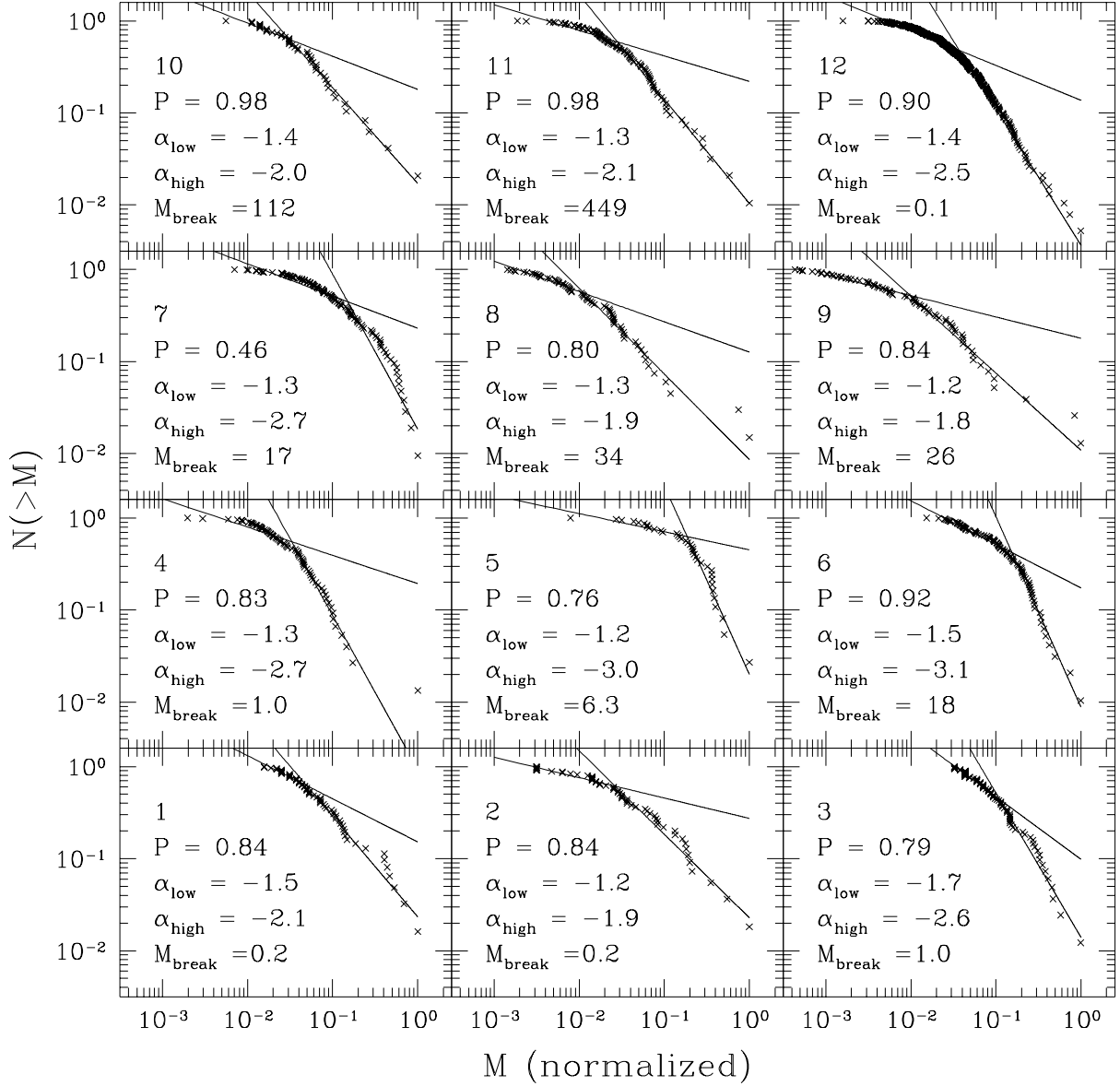}
\caption{Cumulative mass functions (\emph{symbols}) for the 12 data sets 
described in Table~\ref{tab:studies}.  Each CMF is fit by a pure double 
power law (2PL), as described by Equation~\ref{eq:cmf2pl}.  The panel 
labels indicate the data set ID from Table~\ref{tab:studies}, 
goodness-of-fit probability, $P$, and best-fit parameters, $\alpha_{\rm 
low}$, $\alpha_{\rm high}$, and $M_{\rm break}$.  For visual clarity, the 
maximum clump mass in each plot has been normalized to unity; the best fit 
parameters given are for fits to the unnormalized 
data.\label{fig:2plcmfs}}
\end{center}
\end{figure}

\begin{figure}
\begin{center}
\includegraphics[width=\columnwidth]{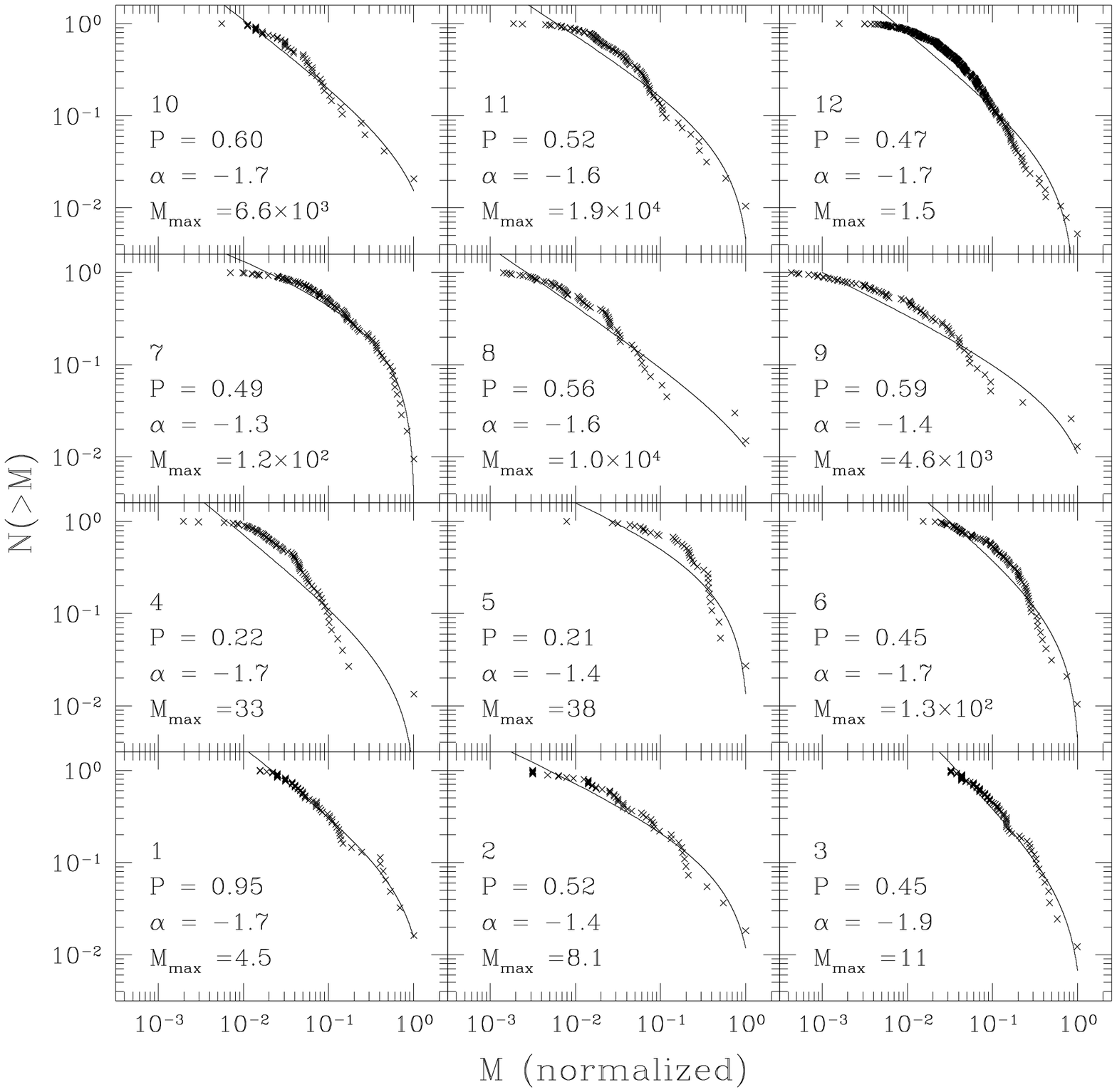}
\caption{Cumulative mass functions (\emph{symbols}) for the 12 data sets 
described in Table~\ref{tab:studies}.  Each CMF is fit by a single power 
law (1PLMM) described by Equation~\ref{eq:cmf1plmm}.  The panel labels indicate 
the data set ID from Table~\ref{tab:studies}, goodness-of-fit probability, 
$P$, and best-fit parameters, $\alpha$ and $M_{\rm max}$.  For visual 
clarity, the maximum clump mass in each plot has been normalized to unity; 
the best fit parameters given are for fits to the unnormalized 
data.\label{fig:1plcmfs}}
\end{center}
\end{figure}

\begin{figure}
\begin{center}
\includegraphics[width=\columnwidth]{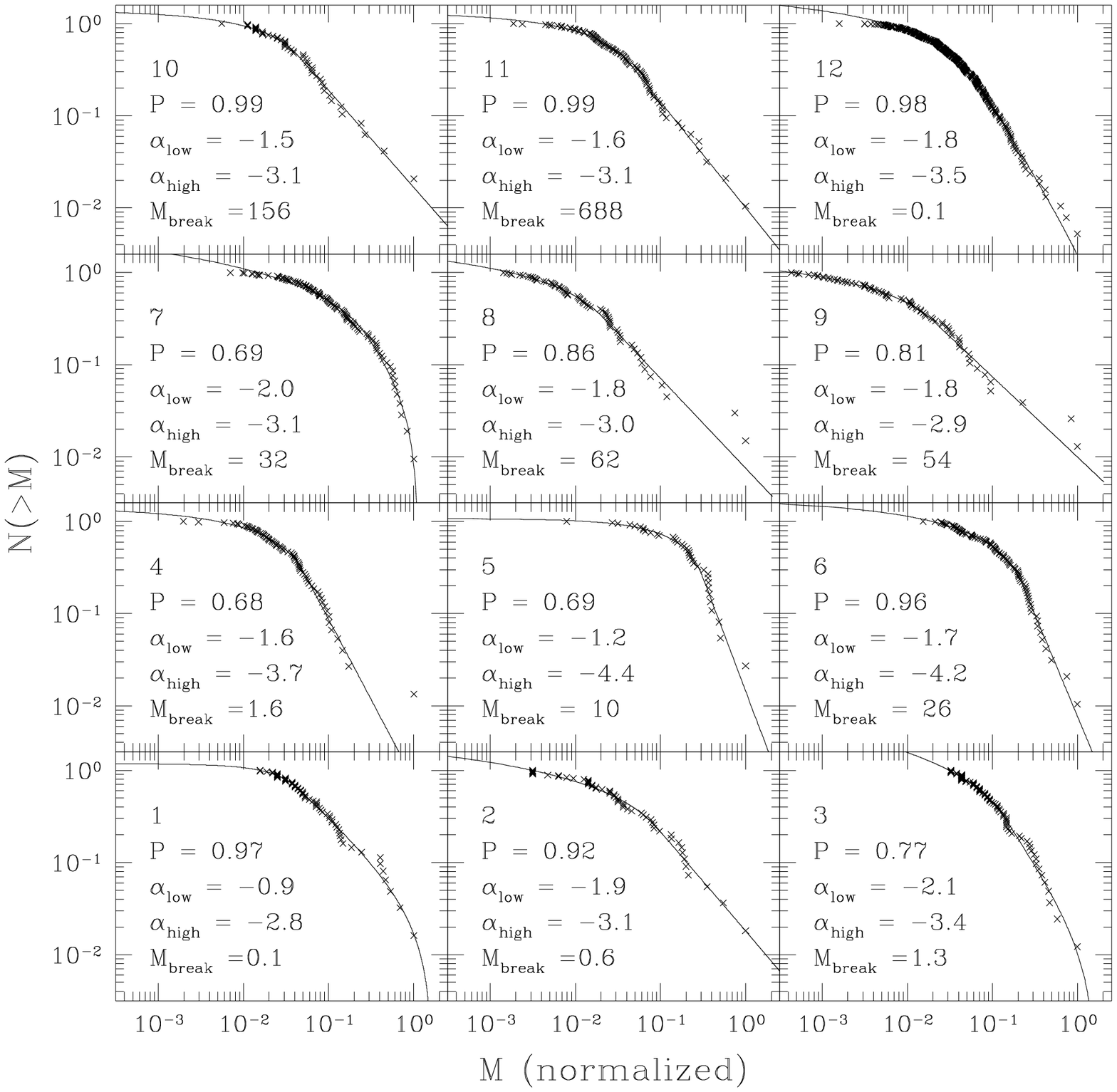}
\caption{Cumulative mass functions (\emph{symbols}) for the 12 data sets 
described in Table~\ref{tab:studies}.  Each CMF is fit by the double power 
law (2PLMM) described by Equation~\ref{eq:cmf2plmm}.  The panel labels indicate 
the data set ID from Table~\ref{tab:studies}, goodness-of-fit probability, 
$P$, and best-fit parameters, $\alpha$ and $M_{\rm max}$.  For visual 
clarity, the maximum clump mass in each plot has been normalized to unity; 
the best fit parameters given are for fits to the unnormalized 
data.\label{fig:2plmmcmfs}}
\end{center}
\end{figure}

\begin{figure}
\begin{center}
\includegraphics[width=\columnwidth]{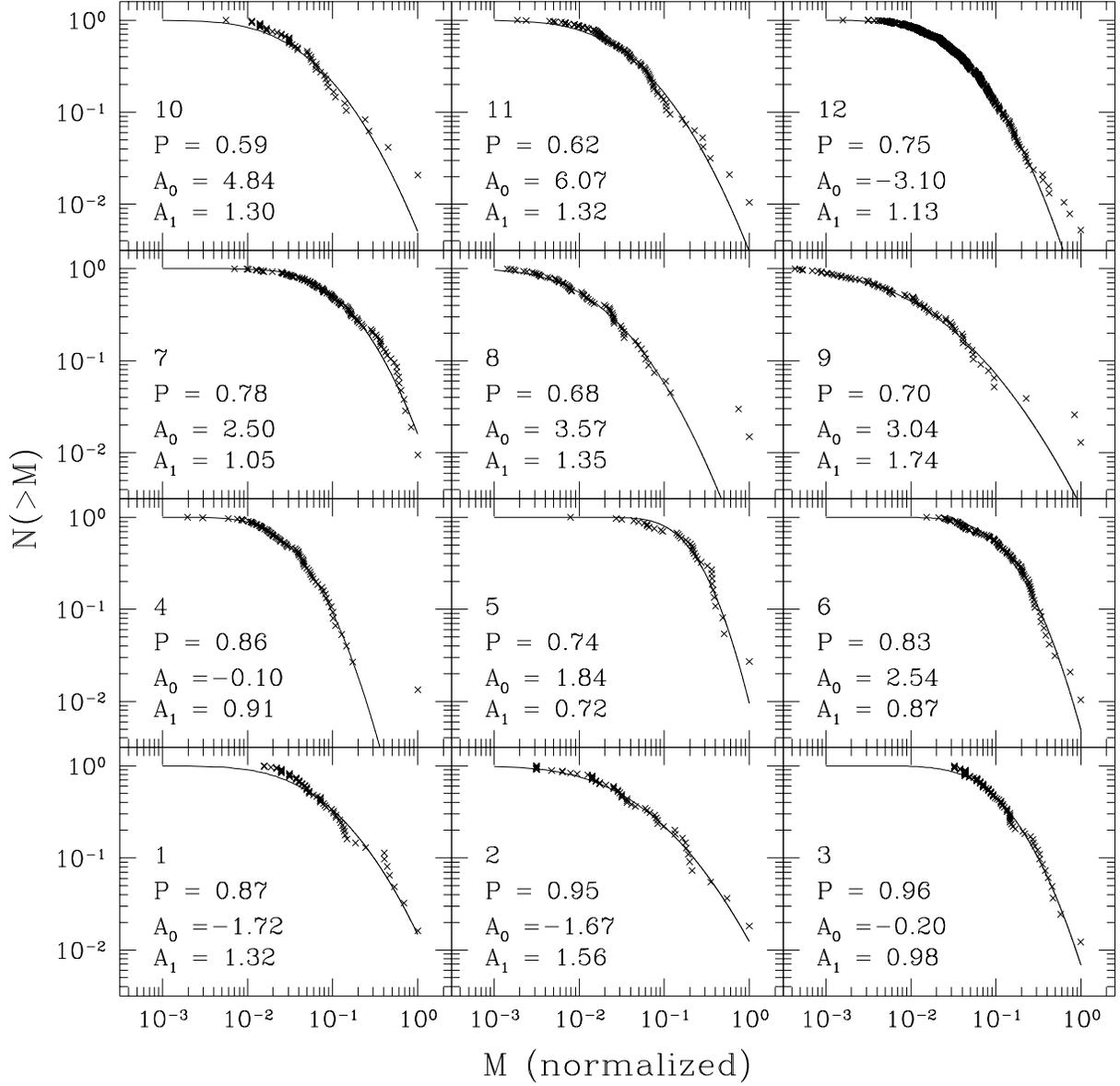}
\caption{Cumulative mass functions (\emph{symbols}) for the 12 data sets
described in Table~\ref{tab:studies}.  Each CMF is fit by a lognormal
clump mass distribution (LOGN), as described by Equation~\ref{eq:cmflogn}.  
The
panel labels indicate the data set ID from Table~\ref{tab:studies},
goodness-of-fit probability, $P$, and best-fit parameters, $A_{0}$ and
$A_{1}$.  For visual clarity, the maximum clump mass in each plot has been
normalized to unity; the best fit parameters given are for fits to the
unnormalized data.\label{fig:logncmfs}}
\end{center}
\end{figure}

	Figures~\ref{fig:2plcmfs}--\ref{fig:logncmfs} show all eleven 
measured CMFs with the best fit versions of these functions: 2PL, 1PLMM, 
2PLMM, and LOGN.  The CMF of the simulated clumps from TP04 is included 
for comparison.  The $P$ values for all of the fits are summarized in 
Table~\ref{tab:fitpvals}.

\begin{deluxetable}{lccccc}
\tablewidth{0pt}
\tablecaption{$P$ Values for Fits to CMFs\label{tab:fitpvals}}
\tablehead{
\colhead{} &
\colhead{} &
\multicolumn{4}{c}{Fit Type} \\
\cline{3-5}
\colhead{Region Name} &
\colhead{Data Set ID} &
\colhead{2PL} & 
\colhead{1PLMM} & 
\colhead{2PLMM} & 
\colhead{LOGN} 
}
\startdata
$\rho$ Oph (1.3 mm) & 1 & 0.84 & 0.95 & {\bf 0.97} & 0.87 \\
$\rho$ Oph (850 $\mu$m) & 2 & 0.84 & 0.52 & 0.92 & {\bf 0.95} \\
Orion B (850 $\mu$m) & 3 & 0.79 & 0.45 & 0.77 & {\bf 0.96} \\
Orion B (850 $\mu$m) & 4 & 0.83 & 0.22 & 0.68 & {\bf 0.86} \\
Lagoon & 5 & {\bf 0.76} & 0.21 & 0.69 & 0.74 \\
M17 (450 $\mu$m) & 6 & 0.92 & 0.45 & {\bf 0.96} & 0.83 \\
M17 (850 $\mu$m) & 7 & 0.46 & 0.49 & 0.69 & {\bf 0.78} \\
NGC 7538 (450 $\mu$m) & 8 & 0.80 & 0.56 & {\bf 0.86} & 0.68 \\
NGC 7538 (850 $\mu$m) & 9 & {\bf 0.84} & 0.59 & 0.81 & 0.70 \\
W43 & 10 & 0.98 & 0.60 & {\bf 0.99} & 0.59 \\
RCW 106 & 11 & 0.98 & 0.52 & {\bf 0.99} & 0.62 \\
Run B5 of TP04 & 12 & 0.90 & 0.47 & {\bf 0.98} &0.75 \\
\enddata
\tablecomments{Fit types are: direct double power-law fit to the CMF 
(2PL), single power law DMF integrated to a finite upper mass limit 
(1PLMM), double power law DMF integrated to a finite upper mass limit 
(2PLMM), and a lognormal DMF (LOGN).  The highest $P$ value for each 
star-forming region is displayed in bold
text.}
\end{deluxetable}

\subsection{Fits to Observed Clump Mass Functions}

	Figures~\ref{fig:2plcmfs}--\ref{fig:logncmfs} and the P values 
shown in Table~\ref{tab:fitpvals} indicate that none of the observational 
CMFs is best fit by a single power law, 7 are best fit by some form of 
the double power law (either 2PL or 2PLMM), and 4 are best fit by a 
lognormal clump mass distribution (LOGN).  Before discussing each case in 
turn, we encourage the reader to keep in mind that small differences in 
the $P$ values between two almost equally good fits are probably not 
meaningful.  In calculating the $P$ values, we had to use the bootstrap 
approximation in place of the frequently-unavailable clump mass 
uncertainties, and this makes the $P$ values themselves uncertain.

\noindent $\boldrho~{\bf Oph}$ --- The \citet{man98} 1.3~mm core CMF (data 
set 1) is best fit by the 2PLMM distribution, although the 1PLMM fit is 
only marginally poorer.  Similarly, the \citet{dj2000b} 850~$\mu$m core 
CMF (set 2) is best fit by the LOGN distribution, but the 2PLMM fit is 
only marginally poorer.  Given that the two data sets represent the same 
clump sample, we expect that the same function should describe them.  
Either the 2PLMM or LOGN functions could serve this purpose, but the 2PLMM 
function gives a slightly higher mean $P$ value.  Interestingly, the LOGN 
fit, which has only 2 parameters, nearly matches the quality of the 2PLMM 
function, which has 5.

\noindent {\bf Orion B} --- In this region the lognormal mass function 
provides the best fit to both measurements of the CMF.  The 2PL function 
provides an almost equally good fit to data set~4, but not to data set~3.  
Visual inspection of Figures~\ref{fig:2plcmfs} and \ref{fig:logncmfs} 
affirms the superiority of the LOGN fit.  If we exclude the massive 
outlier clump at the high-mass end of set~4, the $P$ value for the 
LOGN fit rises to 0.99 and that for the 2PL fit falls to 0.78.  We suspect 
that the massive outlier is probably a composite clump which would resolve 
into multiple less-massive objects if observed at higher angular 
resolution.

\noindent {\bf Lagoon} --- Although \citet{tot02} argued for a single 
power law fit to this mass function (set 5), the $P$ values indicate that 
the 2PL function provides a significantly better fit ($P = 0.76$ versus $P 
= 0.21$).  The lognormal fit, with $P = 0.74$, is only marginally poorer 
than the 2PL fit.  Note that the study from which these data were obtained 
is unique in having used a by-eye clump extraction method.

\noindent {\bf M17} --- As in Paper~II, we find that the 450~$\mu$m CMF 
(set 6) is better fit by a double power law (2PLMM, in this case), while 
the 850~$\mu$m CMF (set 7) is better fit by a lognormal distribution.  
Both the LOGN and 2PLMM functions might be said to provide credible fits 
to both data sets, though again the LOGN function does this using three 
fewer parameters than does the 2PLMM function.

\noindent {\bf NGC~7538} --- In this case, the $P$ values show that some 
form of double power law (2PLMM for set 8 and 2PL for set 9) provides 
the best fit to both measurements of the CMF.  The three most massive 
clumps in each CMF correspond to the IRS 1-3, IRS~9, and IRS~11 regions.  
Each of these regions is already known to be forming more than one star 
and therefore likely contains significant unresolved structure.  For 
example, the IRS~1--3 region contains 3 compact {\sc Hii} regions 
\citep{i77,ww74}.  If we omit these outliers from the analysis, the $P$ 
values for the lognormal fits become 0.97 and 0.69 for data sets 8 and 
9, respectively.  The corresponding $P$ values for the 2PLMM fits become 
0.99 and 0.96.  Therefore, we conclude that, although the LOGN function 
remains a contender, the NGC~7538 CMF is probably best fit by a double 
power law.
	
\noindent {\bf W43 and RCW~106} --- These two massive star-forming regions 
provide the clearest cases of CMFs (sets 10 \& 11) which are best fit by 
double power laws.  In both cases, the 2PLMM fits are favored, with $P = 
0.99$, while the 2PL fits both give $P = 0.98$.  No other function 
provides a comparably good fit.

\noindent {\bf Run B5 of TP04} --- Here again, the 2PLMM function provides 
the best fit, with $P = 0.99$.  Visual inspection of the plot suggests 
that the LOGN function fits equally well, although the $P$ values suggest 
otherwise.

	To summarize, the general trend in the fits is toward lognormal 
mass functions among the low-mass star-forming regions and some form of 
the double power law mass function among the massive star-forming regions.  
The two major sources of uncertainty in this analysis are the choice of 
weights and the use of the bootstrap technique.  Both would be mitigated 
if the original clump mass uncertainties were available.  If we use 
uniform weights in making the fits, instead of the default weights of 
$1/N^{2}$, the LOGN function then provides the best fit in 5 cases, the 
2PLMM in 4, and they tie in 3.  With regard to the use of the bootstrap 
technique, we note that although using the real clump mass uncertainties 
would be preferable, the $P$ values obtained herein using the bootstrap 
technique show the same trend as those obtained in Paper~II using the 
proper uncertainties.  We strongly encourage the reporting of both random 
and systematic clump mass uncertainties in future studies.

	A potential fault with the 2PLMM fits can be seen by comparing the 
best fit parameters from the 2PL fits (Fig.~\ref{fig:2plcmfs}) with those 
from the 2PLMM fits (Fig.~\ref{fig:2plmmcmfs}).  The 2PL $\alpha_{\rm 
high}$ values range around the Salpeter value of $\alpha = 2.35$ but the 
mean 2PLMM value is $\left<{\alpha}_{\rm high,2PLMM}\right> = \left<{\alpha}_{\rm 
high,2PL}\right> -1.0$.  Moreover, the mean ratio of the break masses derived 
from the 2PLMM and 2PL fits is 1.6.  The 2PLMM fits are achieving the 
excellent $P$ values they do by shifting the break mass up and fitting the 
top end of the mass function with a very steep power law---one which is 
not obviously related to the power laws which fit the CO clump or stellar 
mass functions, both of which are significantly shallower.  It is also 
important to note that the 2PLMM fits achieve their excellent $P$ values 
using 5 adjustable parameters, compared to 4 for the 2PL fits and only 2 
for the LOGN fits.  For the remainder of the discussion in this paper, we 
will refer primarily to the power law exponents derived from the 2PL fits, 
which are most suitable for direct comparison to those obtained by other 
authors.

\subsection{Interpreting the Fits}

	Previous studies of the core mass function in regions of low-mass 
star formation have emphasized the agreement between the shapes of the 
core and stellar mass functions (e.g. \citealt{ts98,man98,dj2000b}), when 
both are fit with similar numbers of power laws over similar mass ranges.  
The core mass functions in $\rho$~Oph, Orion~B, and Serpens are all well 
fit by a Salpeter-like power law above about 1~$M_{\odot}$ ($\alpha_{\rm 
high} \simeq -2.35$), flatten out somewhat below about 1~$M_{\odot}$ 
($\alpha_{\rm low} \simeq 1.5$), and peak at about 0.1~$M_{\odot}$.

\begin{figure}
\begin{center}
\includegraphics[width=\columnwidth]{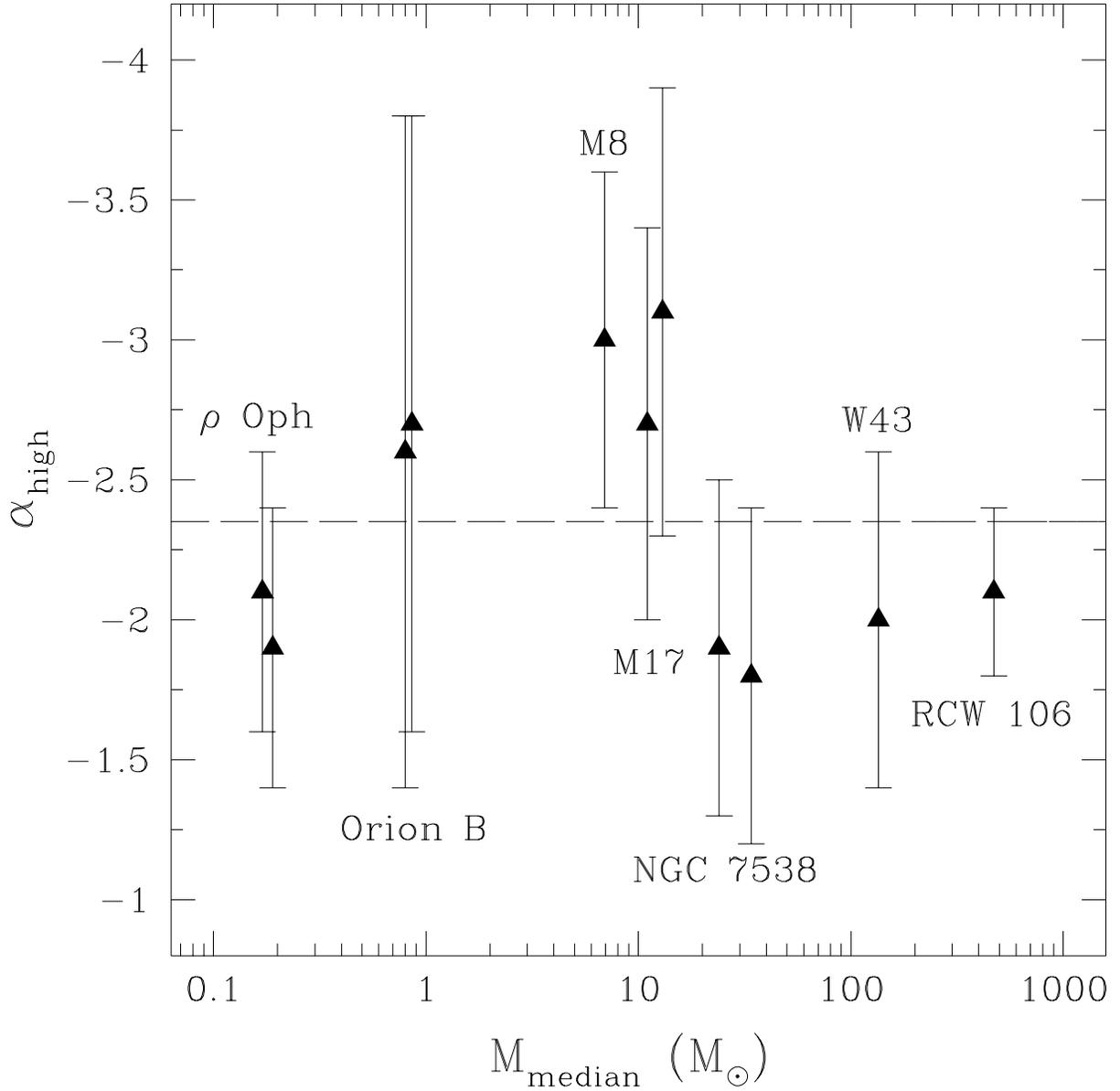}
\caption{Power-law exponent, $\alpha_{\rm high}$, versus median clump mass 
for the 11 millimeter/submillimeter clump mass functions described in 
Table~\ref{tab:studies}.  The $\alpha_{\rm high}$ values are derived from 
the pure double power law fits (2PL, Eq.~\ref{eq:cmf2pl}).  The 
horizontal dashed line indicates the Salpeter exponent, $\alpha_{\rm 
high} = -2.35$ \citep{sal55}. The error bars 
represent the 95\% ($\sim$2$\sigma$) confidence limits from Table~\ref{tab:fitparams}. No trend in $\alpha_{\rm high}$ with mass
is apparent.\label{fig:alphaplot}}
\end{center}
\end{figure}

	In Table~\ref{tab:fitparams}, we summarize the best-fit parameters 
for the 2PL and LOGN fits to the 12 CMFs studied herein.  Fitting all of 
the CMFs in the same way, we find $\alpha_{\rm high}$ values ranging 
between -1.8 and -3.1, comparable to the range of power law exponents 
found in fits to the $\gtrsim 1 M_{\odot}$ stellar IMF \citep{kroupa02}.  
The mean of the $\alpha_{\rm high}$ values is -2.4$\pm$0.1, or 
-2.3$\pm$0.2 if we include only the 7 observational CMFs which are best 
fit by double power laws.  Both measures of the mean $\alpha_{\rm high}$ 
are compatible with the Salpeter IMF, despite the differences in the clump 
mass ranges over which the various $\alpha_{\rm high}$ values were 
derived.  Similarly, if the low-mass ends of all 11 observational CMFs are 
fit with power laws, the resulting exponents, $\alpha_{\rm low}$, range 
between -1.2 and -1.7, with a mean of -1.4$\pm$0.1.  This result also 
accords with the shallowing of the stellar IMF at its low-mass end, 
despite the differences in the clump mass scales over which this 
shallowing occurs.  Perhaps most interesting is that, as shown in 
Figure~\ref{fig:alphaplot}, we find no trend in the $\alpha_{\rm high}$ 
values with the median clump mass in each region.  This result applies 
over nearly four orders of magnitude in median clump mass.  The power law 
indices derived from fits to the mass functions of submillimeter continuum 
clumps/cores offer no clear way to distinguish between low-mass regions, 
such as $\rho$~Oph, and very high-mass regions, such as RCW~106.  The 
median clump mass in RCW~106 is nearly 500~$M_{\odot}$, which suggests 
that most of the clumps are probably cluster-forming objects, yet the 
RCW~106 clump mass function and the $\rho$~Oph core mass function may be 
fit with compatible power laws.

\begin{deluxetable}{lccccccccc}
\tabletypesize{\scriptsize}
\tablewidth{0pt}
\tablecaption{Best Fit Parameters for 2PL and LOGN Fits\label{tab:fitparams}}
\tablehead{
\colhead{} &
\colhead{Data} &
\multicolumn{4}{c}{LOGN Fit Parameters} &
\colhead{} &
\multicolumn{3}{c}{2PL Fit Parameters} \\
\cline{3-6}\cline{8-10}
\colhead{Region} &
\colhead{Set ID} &
\colhead{$A_{0}$} & 
\colhead{$A_{1}$} & 
\colhead{$M_{\rm peak}$\tablenotemark{a}} & 
\colhead{$\langle M \rangle$\tablenotemark{a}} & 
\colhead{} & 
\colhead{$M_{\rm break}$} & 
\colhead{$\alpha_{\rm low}$} & 
\colhead{$\alpha_{\rm high}$} 
}
\startdata
$\rho$ Oph (1.3 mm) & 1  & -1.7$\pm$0.3 & 1.3$\pm$0.3 & 0.03 $\pm$ 0.01 & 0.43 $\pm$ 0.09 &  & 0.2 $\pm$ 0.2 &  -1.5$\pm$0.3 &  -2.1$\pm$0.5 \\
$\rho$ Oph (850 $\mu$m) & 2  & -1.7$\pm$0.4 & 1.6$\pm$0.4 & 0.016 $\pm$ 0.008 & 0.6 $\pm$ 0.2 &  & 0.2 $\pm$ 0.1 &  -1.2$\pm$0.1 &  -1.9$\pm$0.5 \\
Orion B (850 $\mu$m) & 3  & -0.2$\pm$0.2 & 1.0$\pm$0.2 & 0.31 $\pm$ 0.05 & 1.3 $\pm$ 0.2 &  & 1 $\pm$ 1 &  -1.7$\pm$0.2 &  -2.6$\pm$1.2 \\
Orion B (850 $\mu$m) & 4  & -0.1$\pm$0.2 & 0.9$\pm$0.2 & 0.39 $\pm$ 0.07 & 1.4 $\pm$ 0.2 &  & 1.0 $\pm$ 0.6 &  -1.3$\pm$0.2 &  -2.7$\pm$1.1 \\
Lagoon & 5  & 1.8$\pm$0.3 & 0.7$\pm$0.3 & 3.7 $\pm$ 0.8 & 8 $\pm$ 1 &  & 6 $\pm$ 2 &  -1.2$\pm$0.2 &  -3.0$\pm$0.6 \\
M17 (450 $\mu$m) & 6  & 2.5$\pm$0.2 & 0.9$\pm$0.2 & 6 $\pm$ 1 & 18 $\pm$ 2 &  & 18 $\pm$ 8 &  -1.5$\pm$0.1 &  -3.1$\pm$0.8 \\
M17 (850 $\mu$m) & 7  & 2.5$\pm$0.3 & 1.0$\pm$0.1 & 4.1 $\pm$ 0.7 & 21 $\pm$ 3 &  & 17 $\pm$ 10 &  -1.3$\pm$0.1 &  -2.7$\pm$0.7 \\
NGC 7538 (450 $\mu$m) & 8  & 3.6$\pm$0.3 & 1.3$\pm$0.6 & 6 $\pm$ 3 & 90 $\pm$ 30 &  & 30 $\pm$ 30 &  -1.3$\pm$0.2 &  -1.9$\pm$0.6 \\
NGC 7538 (850 $\mu$m) & 9  & 3.0$\pm$0.4 & 1.7$\pm$0.6 & 1.0 $\pm$ 0.7 & 90 $\pm$ 40 &  & 30 $\pm$ 30 &  -1.2$\pm$0.1 &  -1.8$\pm$0.6 \\
W43 & 10  & 4.8$\pm$0.3 & 1.3$\pm$0.5 & 20 $\pm$ 10 & 300 $\pm$ 90 &  & 110 $\pm$ 80 &  -1.4$\pm$0.2 &  -2.0$\pm$0.6 \\
RCW 106 & 11  & 6.1$\pm$0.2 & 1.3$\pm$0.4 & 80 $\pm$ 30 & 1000 $\pm$ 200 &  & 400 $\pm$ 300 &  -1.3$\pm$0.2 &  -2.1$\pm$0.3 \\
Run B5 of TP04 & 12  & -3.1$\pm$0.1 & 1.1$\pm$0.1 & 0.013 $\pm$ 0.002 & 0.085 $\pm$ 0.007 &  & 0.06 $\pm$ 0.02 &  -1.4$\pm$0.1 &  -2.5$\pm$0.3 \\ \hline
Mean values\tablenotemark{b} & & & {\bf 1.2$\pm$0.1} & & & & & {\bf -1.4$\pm$0.1} & {\bf -2.4$\pm$0.1} \\
\enddata
\tablecomments{All quoted uncertainties correspond to a 95\% 
($\sim$2$\sigma$) confidence interval.}
\tablenotetext{a}{The peak mass, $M_{\rm peak}$, and mean mass, $\langle
M\rangle$, of the lognormal distribution are not parameters of the fit;  
they are calculated from $A_{0}$ and $A_{1}$ using the equations $\langle 
M\rangle = \exp(A_{0} + \frac{1}{2}A_{1}^{2})$ and
$M_{\rm peak} = \exp(A_{0} - A_{1}^{2})$.}
\tablenotetext{b}{Mean values include the 11 observational CMFs but 
exclude the CMF from the simulations of TP04.}
\end{deluxetable}

	We emphasize four key results from this fitting exercise.  First, 
there is the tendency toward lognormal CMFs in less massive star-forming 
regions and power-law CMFs in massive star-forming regions.  Second, there 
is the agreement among the $\alpha_{\rm high}$ and $\alpha_{\rm low}$ 
values from region to region, despite the significant differences in the 
relevant clump mass scales.  Third, there is the presence of a break 
between the steep and shallow sections of the mass function in high-mass 
star-forming regions.  If this break is real, it implies that the clump 
mass function is \emph{not} a single unbroken power law above 
1~$M_{\odot}$.  This result holds true despite the uncertainties in the 
absolute mass calibration in each study (see \S\ref{sec:massunc})  
Finally, there is the agreement between the power law exponents of the 
clump and stellar mass functions, again despite the differences in mass 
scales.

\subsubsection{Lognormal vs. Double Power-Law CMFs}

	There is already ample theoretical evidence that lognormal mass 
functions can result when any sufficiently stochastic process, such as 
turbulent fragmentation, governs the evolution of the clump masses.  
Moreover, lognormal mass functions can be viewed as one extremum of a 
continuum of shapes whose other extremum is the power-law mass function.  
Numerical simulations have repeatedly shown that turbulent fragmentation 
can naturally produce both power-law and lognormal mass functions whose 
shapes are largely independent of the intrinsic mass scale 
\citep{kbb98,kb00,klessen01,pn02,gammie03,tilley04,cb06}.  Several authors 
have invoked the Central Limit Theorem to demonstrate analytically that 
lognormal mass functions arise whenever a large number of random 
parameters act to set the clump masses \citep{larson73,z84,af96}.  
Conversely, power-law mass functions can arise when fewer random 
parameters affect each clump's mass.  These random parameters could derive 
from successive turbulent fragmentations or N-body gravitational 
encounters among clumps.  We might therefore imagine the following 
scenario to explain the trend toward lognormal mass functions in low-mass 
regions and power-law mass functions in high-mass regions (see 
Table~\ref{tab:fitpvals}).  A molecular cloud undergoes turbulent 
fragmentation, producing an initially power-law clump mass function; as 
the clumps continue to fragment and interact, their mass function 
approaches a lognormal distribution; by the time the ``clumps'' have 
become ``cores'', they have achieved a fully lognormal mass distribution.  
As the objects furthest down the chains of turbulent fragmentation and 
N-body interactions, the masses of the cores would have been maximally 
influenced by random processes.

	In this scenario, spatially unresolved observations of a set of 
cores whose mass distribution is lognormal could effectively restore the 
less-evolved double power law mass function by blurring together smaller 
cores into semblances of their antecedent clumps.  This hypothesis 
would explain the trend toward double power law mass functions seen in the 
more distant regions, such as NGC~7538, W43, and RCW~106, where the 
spatial resolution of the observations isn't sufficient to adequately 
distinguish all of the cores from one another.

\subsubsection{The Clump CMF vs. the Stellar IMF}

	If the core/clump mass function in both low- and high-mass 
star-forming regions were found to be consistent with the stellar IMF 
(uncorrected for unresolved multiples), it would relax the need to invoke 
processes such as coalescence and competitive accretion 
\citep{bon97,bon01,bon04} to explain the formation of more massive stars.  
Such consistency would also argue in favor of theories which predict that 
massive stars can form by a scaled-up version of molecular cloud core 
collapse with disk accretion \citep{mt02,mt03,krum05,krum06}.

	As shown in Table~\ref{tab:fitparams}, the mean power law 
exponents from the double power law fits to the 11 observational CMFs are 
$\alpha_{\rm high} = -2.4 \pm 0.1$ and $\alpha_{\rm low} = -1.4 \pm 0.1$.  
\citet{kroupa02} cites power law exponents for the mean Galactic field 
single-star IMF of $-2.3 \pm 0.3$ for $M_{\rm star}/M_{\odot} \geq 0.5$ 
and $-1.3 \pm 0.5$ for $0.08 \leq M_{\rm star}/M_{\odot} < 0.5$.  The 
agreement between the $\alpha_{\rm high}$ values of the clump and stellar 
mass functions is intriguing.  Unlike the $\alpha_{\rm low}$ values, which 
may be strongly affected by incompleteness in the clump mass functions 
(see next section), the $\alpha_{\rm high}$ values should be quite 
reliable.  Thus, the agreement between the stellar and clump $\alpha_{\rm 
high}$ values reinforces the ideas that the stellar IMF originates in the 
mass function of clumps in a fragmented molecular cloud and that the 
stellar mass function can be obtained by an essentially one-to-one 
conversion of clumps into stars (or binaries).  If the high-mass end of 
the stellar IMF can be obtained by converting clumps into stars on a 
more-or-less one-to-one basis, there may be no need to invoke processes 
such as coalescence or competitive accretion.

	\citet{kroupa02} has shown that the variation in the measured 
power law exponents for the stellar IMF is compatible with normal 
measurement variation and need not represent physical region-to-region 
variation.  We suggest that the same description could apply equally well 
to the range of power law exponents derived from clump mass functions in 
different regions (see Fig.~\ref{fig:alphaplot}).  More measurements of 
$\alpha_{\rm high}$ would be required to confirm this hypothesis in the 
cases of both the stellar IMF and the clump mass function.  
\citet{elmegreen04} has shown that real region-to-region variations in the 
IMF are physically plausible, but may be masked by small-number 
statistics, which is certainly a concern with the clump/core mass 
functions taken to date.

	Can we identify a mass scale at which the apparent agreement 
between the clump and stellar mass functions breaks down?  Of the 
parameters which control the shapes of the fitted functions, namely 
$A_{1}$ for the lognormal fits and the $\alpha$ values for the power law 
fits, none shows a trend with the median clump mass of the regions (see 
Table~\ref{tab:fitparams}).  Thus, the fits suggest that the shape of the 
clump mass function is not very sensitive to the clump mass range being 
fit.  There is also no clear break between the CMFs which do mirror the 
shape of the stellar IMF and those which do not, despite the breadth in 
the clump mass ranges.  Again, this surprising result suggests that the 
intrinsic shape of the stellar IMF may be apparent in the clump mass 
function on mass and linear scales substantially larger than those 
previously considered.  

\subsubsection{Incompleteness and Spatial Filtering}
	
	Both interferometry and the chopping technique used by single-dish 
submillimeter telescopes filter out emission on some spatial scales.  
Clump-finding algorithms may have a similar effect.  Therefore, all of the 
observations used to generate the CMFs discussed in this paper are subject 
to some form of spatial filtering, and hence to incompleteness and 
uncertainty.  We are therefore motivated to ask whether spatial filtering 
and related effects introduced during data acquisition and reduction can 
introduce a sufficient number of random perturbations to the clump masses 
to ensure that a lognormal mass function is obtained.

	Spatial filtering of flux can systematically alter the derived 
clump masses.  The magnitude of the discrepancy would depend on the size 
(and therefore the mass) of each clump.  Although this possibility should 
be explored further, several arguments suggest it does not substantially 
alter the shape of the clump mass function.  First, the mass function of 
clumps extracted from infrared dust extinction maps, which are essentially 
free of spatial filtering, is very similar to those derived from 
submillimeter continuum observations (J. Alves, private communication). If 
the clump extraction algorithms exerted a strong spatial filtering effect, 
we would need to understand how at least three different methods used by 
the authors of the papers studied here all produce similar mass functions. 
Moreover, the effects of spatial filtering would affect observations of 
both low- and high-mass star-forming regions.  If it were a significant 
effect, it seems unlikely that observations of low-mass star-forming 
regions would still manage to produce a core mass function which appears 
to agree with the stellar IMF.  Finally, we note that the clump CMF 
extracted from the turbulent fragmentation simulations of TP04 is not 
subject to observational spatial filtering and its shape is entirely 
compatible with those of the observational CMFs (TP04 used a custom clump 
extraction technique).  For these reasons, it is difficult to 
explain how the shape of the clump mass function, and its similarity from 
region to region might be accounted for by spatial filtering.

	Whether or not the clump mass functions are strongly affected by 
spatial filtering, they are certainly all affected by incompleteness.  
For spatially extended objects, the detection threshold is a limiting 
surface brightness, meaning that a clump of any mass could go undetected 
if its flux were spread thinly across the sky.  Thus, the clump sample may 
be incomplete in any mass range.  At the moment, there is no reliable way 
to correct millimeter and submillimeter continuum clump/core mass 
functions for incompleteness.  If incompleteness is a significant 
determinant of the shape of each CMF, its effects are likely to be felt 
most strongly at their low-mass ends, where it may be partly responsible 
for the shallower slope of the mass function.  Consequently, the degree of 
incompleteness might influence the position of the break, $M_{\rm break}$, 
between the two power laws when a double power law is fitted to a given 
mass function.  In most studies to date, the distribution of clumps in 
mass-radius space shows a (sometimes slight) separation from the detection 
threshold at its high-mass end \citep{paperi,paperii,m01}.  Hence, we do 
not think it likely that the fitted $\alpha_{\rm high}$ values are 
strongly affected by sensitivity-induced incompleteness.  The $\alpha_{\rm 
low}$ values, however, are likely to be significantly impacted by 
incompleteness.

	There are several ways to test for the related effects of 
incompleteness and spatial filtering.  High spatial resolution 
observations of the more distant star-forming regions in the sample will 
reveal the extent to which currently unresolvable structure will change 
the shape of the clump/core mass function.  Observations of more distant 
regions should also mitigate the effects of chopping by limiting the loss 
of emission to spatial scales much larger than those of typical molecular 
cloud cores.  Comparison of submillimeter continuum and dust extinction 
maps of identical regions should provide strong constraints on the effect 
of spatial filtering in the former type of observations.  Finally, 
observations with total power detectors, such as SCUBA-2, will eliminate 
the effects of chop-based spatial filtering.  SCUBA-2's greater 
sensitivity will also help reveal those clumps not detectable by the 
present generation of submillimeter and millimeter detectors.

\subsubsection{Uncertainties in Absolute Mass Calibration}
\label{sec:massunc}

	It should be noted that, while there are significant uncertainties 
in the absolute calibration of clump masses, these uncertainties are not 
large enough to fundamentally change the conclusions of the preceding 
analysis.  The calibration of the absolute mass scale in each study we 
reference relies on assumptions about the dust temperature and emissivity 
in each region.  Typically, as in Papers~I and II, values typical of each 
region are applied to all of the clumps it contains.  The dust emissivity 
in star-forming regions may be uncertain by up to a factor of 2 (i.e. it 
may range between $\sim$1--2) and the temperature may be uncertain by 
perhaps a factor of four (i.e. the mean clump temperature is likely to lie 
within 10--40~K).  In the most extreme case, whereby the mean temperature 
and dust emissivity in a region are assumed to be 10~K and 1, 
respectively, but are actually 40~K and 2, the clump masses would all be 
overestimated by a factor of $\sim$30.  However, referring to 
Table~\ref{tab:fitparams}, we see that even this large uncertainty is 
insufficient to account for the observed variation in the peak mass, 
$M_{\rm peak}$ from the LOGN fits or the break mass, $M_{\rm break}$, from 
the 2PL fits.  Both of these parameters range over $> 3$ orders of 
magnitude, which is a much greater variation than can be accounted for by 
the largest plausible uncertainties in the dust emissivity and 
temperature.  In addition, the mass calibrations are unlikely to be 
incorrect by the same amount or in the same direction in all of the 
studies.  Finally, we emphasize that uncertainties in the absolute mass 
calibration do not affect the \emph{shapes} of the various clump mass 
functions.

\section{REGION-TO-REGION VARIATIONS IN THE CLUMP MASS FUNCTION}
\label{sec:compare}

	The clump mass function is one of the most important descriptors 
of the initial conditions for star formation.  It tells us how the masses 
of pre-stellar condensations are distributed, leaving us to determine how 
such a distribution could give rise to the observed distribution of 
stellar masses.  Thus, it is important to understand how the clump mass 
function varies from region to region, between low- and high-mass 
star-forming regions, and whether it varies more or less than the stellar 
IMF.  This information will help evaluate theories which posit different 
formation mechanisms for stars of different masses.  For example, if 
low- and high-mass star forming regions had systematically different clump 
mass functions but compatible stellar mass functions, it would be evidence 
that the mechanisms of low- and high-mass stars were different.

\begin{figure} 
\begin{center} 
\includegraphics[width=6.0in]{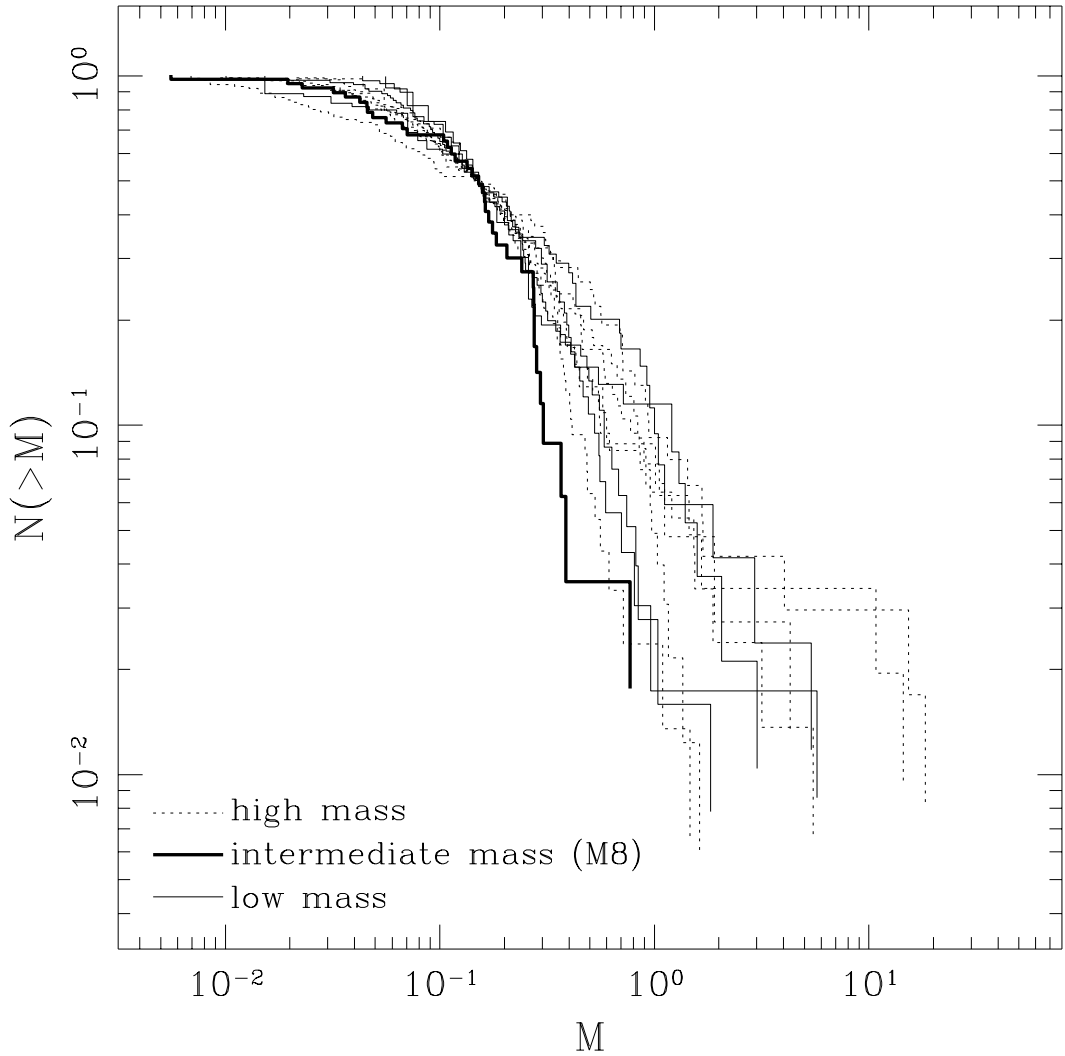}
\caption{Cumulative clump mass functions for all 11 observational studies 
listed in Table~\ref{tab:studies}.  The masses of the clumps in each CMF 
have been rescaled such that all clump sets have a common median mass.  
CMFs of low-mass star-forming regions ($\rho$~Oph and Orion~B) are plotted 
with solid lines, that of M8 with a thick solid line, and those of the 
massive star-forming regions (M17, NGC~7538, W43, and RCW~106) with dotted 
lines.  The new mass scale is arbitrary.  \label{fig:allcmfs}}
\end{center} 
\end{figure}

	In Figure~\ref{fig:allcmfs}, we plot all 11 of the observational 
CMFs from Table~\ref{tab:studies} in the same panel.  To facilitate 
comparisions among the mass functions, they have all been renormalized such 
that they share a common median clump mass.  This renormalization 
preserves the shape of each mass function.  Like 
Figure~\ref{fig:alphaplot}, Figure~\ref{fig:allcmfs} shows that there is 
no clear trend in the shape of the mass function with the median mass in 
each region.  Particularly at their high-mass ends, where incompleteness 
is least significant, there is no clear segregation between the CMFs of 
low- and high-mass star-forming regions.

\begin{figure} 
\begin{center} 
\includegraphics[width=5.0in]{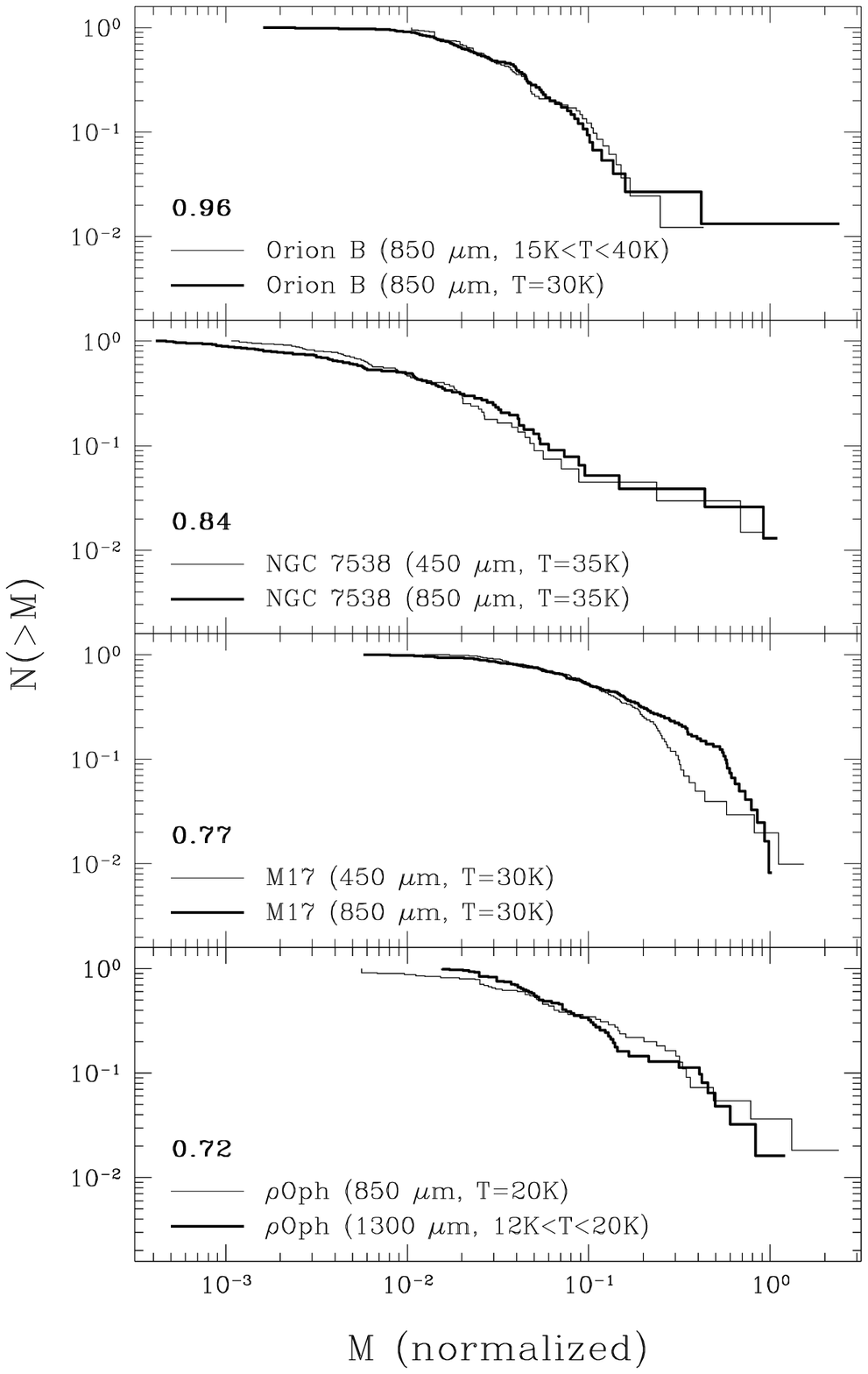}
\caption{Comparison of the cumulative clump mass functions in four
different star-forming regions, using data at different wavelengths
and/or drawn from different studies.  In each panel, the bold number 
indicates the K-S test $P$ value for the comparison of the two CMFs.
\label{fig:same_mfs}} 
\end{center} 
\end{figure}

	In the previous section, we described how parameters derived from 
fits to the clump mass function vary from one star-forming region to 
another.  However, to more precisely quantify the extent to which the 
shape of the clump mass function varies from region to region, it is best 
to use non-parametric techniques.  Such techniques do not rely on prior 
knowledge of the functional form of the clump mass function.  One such 
test is the \ks\ (K-S) test, which measures the probability , $P$, that 
two sets of clump masses represent random samplings from the same parent 
distribution.  We compared each of the mass functions listed in 
Table~\ref{tab:studies} to the others using the K-S test.  To establish a 
baseline for interpreting the K-S test $P$ values, we compare similar sets 
of mass functions.  In Figure~\ref{fig:same_mfs}, we plot pairs of 
independent measurements of the CMF in each of the four regions where 
multiple measurements are available ($\rho$~Oph, Orion~B, NGC~7538, and 
M17).  The $P$ values for these comparisons range from 0.72 for $\rho$~Oph 
to 0.96 for Orion~B.  The lowest $P$ value results from comparing studies 
\citep{man98,dj2000b} of $\rho$~Oph which differ in their authors, 
continuum wavebands, clump extraction techniques, and assumptions about 
the clump temperatures.  The two intermediate $P$ values come from our own 
studies of NGC~7538 (Paper~I) and M17 (Paper~II).  Each pair differs only 
by the continuum waveband of the observations.  The highest $P$ value 
derives from the comparison of the two Orion~B studies \citep{m01,dj2001}, 
and they differ by authors, clump extraction techniques, and assumptions 
about temperatures, but not in continuum waveband.  In principle, each 
pair of mass functions should be the nearly identical, as they represent 
different measures of essentially the same thing.  Thus, the magnitude of 
the difference between the two should primarily reflect differences 
introduced during data reduction and analysis.  Hence, we set our 
threshold for statistically significant similarity between two CMFs at $P 
= 0.72$, the lowest measured probability for comparisons of similar data 
sets.  Note that these results agree well with those from the previous 
section: we find the highest K-S $P$ value when comparing the two Orion~B 
studies, which were both found to be well-fit by lognormal distributions 
in \S\ref{sec:funcform}, and the lowest K-S $P$ value when comparing the 
$\rho$~Oph mass functions, which were found to be best-fit by different 
functions.

\begin{figure} 
\begin{center} 
\includegraphics[width=\columnwidth]{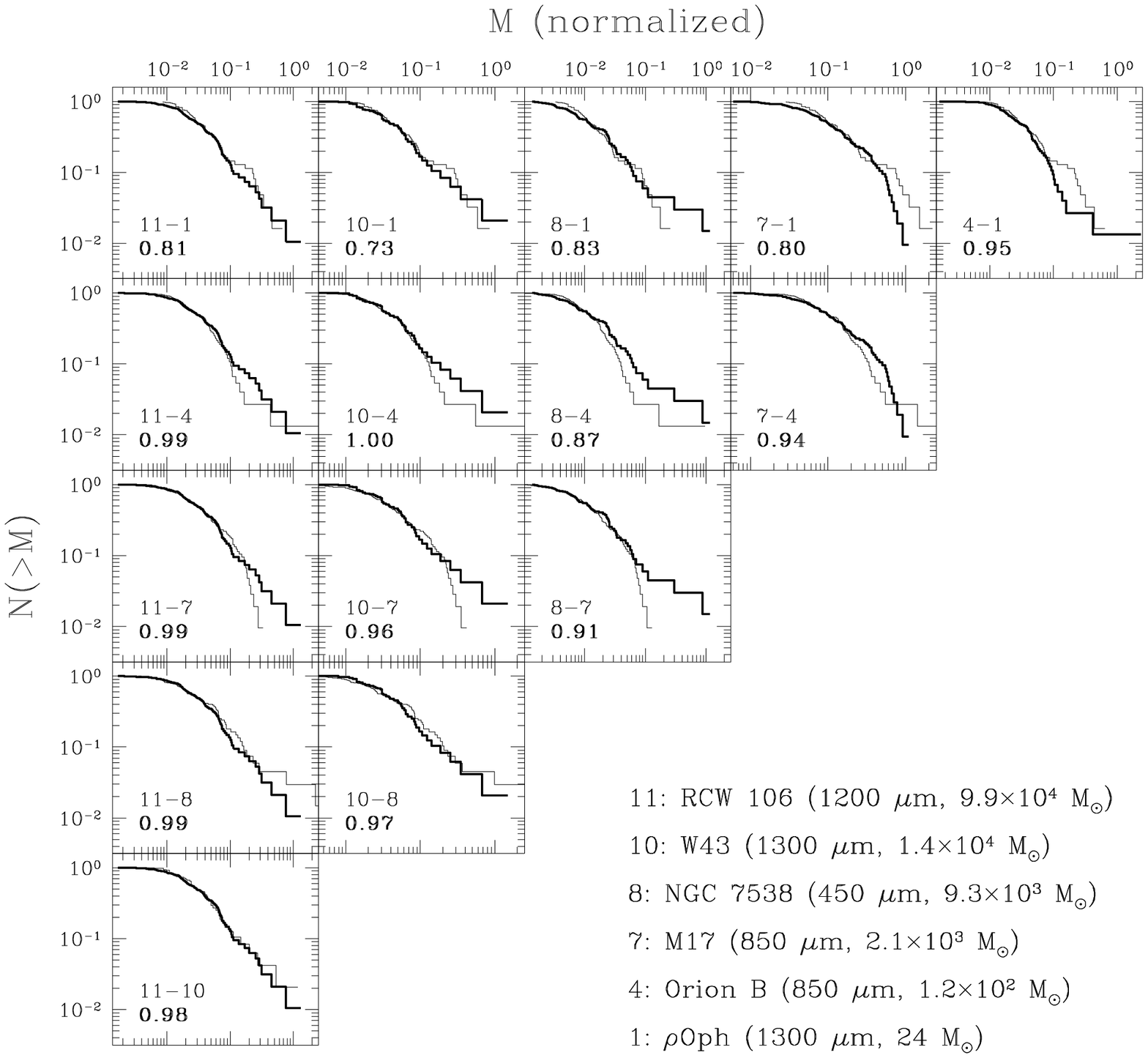}
\caption{Comparison of the cumulative mass functions (CMFs) of six
different star-forming regions, spanning more than five orders of
magnitude in clump mass (0.05 to $\sim$16000 M$_{\odot}$).  In the lower
left corner of each panel, the regions being compared are identified by
their ID numbers, $x$--$y$, from Table~\ref{tab:studies}. The most massive
clump in CMF $x$ (\emph{thick line}) is normalized to mass unity and CMF
$y$ (\emph{thin line}) is normalized accordingly (see text).  The boldface
number in each panel is the \ks\ $P$ value for the comparison (see text 
for
interpretation).  For each region, the wavelength at which the continuum
observations were made and the total mass of clumps detected are given in
the legend. \label{fig:ksmulti}}
\end{center} 
\end{figure}

	Table~\ref{tab:allps} shows the $P$ values for pair-wise K-S test 
comparisons of all of the mass functions described in 
Table~\ref{tab:studies}.  Clearly, a direct K-S comparison of regions such 
as $\rho$ Oph and RCW~106 would yield a $P$ value of 0 because their clump 
mass ranges do not overlap.  To enable a comparison of the \emph{shapes} 
of any two mass functions regardless of their intrinsic mass scales, we 
renormalize each pair of mass functions so they share a common median 
clump mass, as we did for all of the CMFs in Figure~\ref{fig:allcmfs}.  
In Figure~\ref{fig:ksmulti} we plot the pair-wise comparisons of a 
representative set of 6 clump mass functions along with the $P$ value for 
each comparison.  The $P$ values for the 15 unique comparisons range from 
0.73 to 1.00, implying that, in every case, when the mass functions are 
set on a common mass scale, there is a significant likelihood that they 
represent random samples from the same clump mass distribution.  
Examining the $P$ values for all 55 unique inter-comparisons between the 
11 observational CMFs (Table~\ref{tab:allps}), we find that only 11 fall 
below the threshold of statistical significance ($P=0.72$), though none is 
lower than $P=0.56$.  Of these 11, 4 are associated with the \citet{m01} 
study of Orion~B, which differs from the rest in having large numbers of 
duplicate clump masses.  Long runs of identical masses open gaps between 
the \citet{m01} CMF and the others.  The K-S test interprets these gaps as 
statistically significant discrepancies when, in fact, they may only 
reflect the number of significant figures in the reported clump masses.  
This effect may account for the lower average $P$ values for comparisons 
involving the \citet{m01} Orion~B CMF, compared to those involving the 
very similar study of Orion~B by \citet{dj2001}, whose $P$ values indicate 
statistically significant similarity in every case.

\begin{deluxetable}{lcccccccccccc}
\tabletypesize{\scriptsize}
\tablewidth{0pt}
\tablecaption{\ks\ $P$ Values for Mass Function 
Comparisons\label{tab:allps}}
\tablehead{
\colhead{Region Name} &
\colhead{Data Set ID} &
\colhead{2} & 
\colhead{3} & 
\colhead{4} & 
\colhead{5} & 
\colhead{6} & 
\colhead{7} & 
\colhead{8} & 
\colhead{9} & 
\colhead{10} & 
\colhead{11} & 
\colhead{12} 
}
\startdata
$\rho$ Oph (1.3 mm) & 1       &  0.72   &  0.87   &  0.95   &  0.82   &  {\bf 0.63}   &  0.80   &  0.83   &  {\bf 0.60}   &  0.73   &  0.81   &  {\bf 0.68}    \\
$\rho$ Oph (\eight) & 2       & \nodata &  0.77   &  0.95   &  {\bf 0.65}   &  0.81   &  1.00   &  0.98   &  0.77   &  0.90   &  0.96   &  0.96    \\
Orion B (\eight)    & 3       & \nodata & \nodata &  0.96   &  {\bf 0.67}   &  {\bf 0.68}   &  {\bf 0.64}   &  0.83   &  {\bf 0.60}   &  0.76   &  0.77   &  {\bf 0.69}    \\
Orion B (\eight)    & 4       & \nodata & \nodata & \nodata &  0.92   &  0.91   &  0.94   &  0.87   &  0.79   &  1.00   &  0.99   &  0.98    \\
Lagoon              & 5       & \nodata & \nodata & \nodata & \nodata &  0.90   &  {\bf 0.70}   &  {\bf 0.56}   &  0.72   &  0.91   &  {\bf 0.64}   &  0.70    \\
M17 (\four)         & 6       & \nodata & \nodata & \nodata & \nodata & \nodata &  0.77   &  0.88   &  0.82   &  0.98   &  0.99   &  0.81    \\
M17 (\eight)        & 7       & \nodata & \nodata & \nodata & \nodata & \nodata & \nodata &  0.91   &  0.78   &  0.96   &  0.99   &  0.99    \\
NGC 7538 (\four)    & 8       & \nodata & \nodata & \nodata & \nodata & \nodata & \nodata & \nodata &  0.84   &  0.97   &  0.99   &  0.95    \\
NGC 7538 (\eight)   & 9       & \nodata & \nodata & \nodata & \nodata & \nodata & \nodata & \nodata & \nodata &  {\bf 0.71}   &  0.87   &  0.81    \\
W43                 & 10      & \nodata & \nodata & \nodata & \nodata & \nodata & \nodata & \nodata & \nodata & \nodata &  0.98   &  0.97    \\
RCW 106             & 11      & \nodata & \nodata & \nodata & \nodata & \nodata & \nodata & \nodata & \nodata & \nodata & \nodata &  0.98    \\
\enddata
\tablecomments{Bold text indicates those values which fall below the
threshold of statistical significance ($P=0.72$; see text).}
\end{deluxetable}

	Thus, without reference to the functional form of the clump/core 
mass function, we find that, when set on a common mass scale, all of the 
CMFs have a statistically significant likelihood of representing random 
samples drawn from the same parent distribution.  In other words, the CMFs 
all appear to have the same (or very similar) \emph{shapes}, despite the 
differences in their intrinsic mass \emph{scales}.  Because the K-S test 
is least sensitive to the high- and low-mass tails of the CMFs, it is less 
affected by the few most massive clumps in each region which, as we saw in 
\S\ref{sec:funcform}, can bias the determination of the functional form of 
the CMF.  The combined results of the parametric fits from 
\S\ref{sec:funcform} with the non-parametric K-S test comparisons provide 
compelling evidence that the shapes of the clump and core mass 
distributions in all of the regions considered are strongly linked to each 
other and perhaps also to the stellar IMF.

\section{SUMMARY}

	In this paper, we have examined the issue of the functional form 
of the clump/core mass function derived from millimeter/submillimeter 
continuum observations of Galactic star-forming regions.  We have 
demonstrated that, even in the case of a strict single power-law mass 
function, great care must be taken in fitting the cumulative form of the 
mass function to accurately measure the exponent of this power law.  
Unexpected curvature can be introduced into the CMF when the range of 
masses or the number of objects is small, particularly when the power law 
is shallow ($\Delta N/\Delta M \propto M^{\alpha}$ with $\alpha \gtrsim 
-2$).  In some cases, an intrinsically single power-law mass function 
might appear better fit by two or more power laws.  In practice, the mass 
function is unlikely to be a pure single power law, so the potential to be 
mislead when fitting the cumulative mass function is even greater.

        We have fit 11 independent measurements of the clump/core mass 
function extracted from millimeter and submillimeter continuum 
observations of 7 star-forming regions.  The masses of the clumps/cores in 
these regions collectively span more than five orders of magnitude.  We 
find that, in regions where the median clump/core mass is less than a few 
$M_{\odot}$, the clump/core mass function is best fit by a lognormal 
distribution.  In more massive regions, a double power law provides a 
better fit to the clump CMF.  At intermediate masses, both the lognormal 
and the double power law adequately describe the CMF.  This difference in 
the functional form of the clump mass function may represent an evolution 
toward a lognormal mass function at lower masses and smaller spatial 
scales.  This interpretation is consistent with theoretical explanations 
for the origin of a lognormal mass function via stochastic processes such 
as turbulent fragmentation and gravitational N-body interactions.

	The shape of the clump/core mass function does not appear to be a 
strong function of the intrinsic mass scale.  The transition from 
lognormal to double power-law mass functions is fairly subtle.  The mean 
value of the power-law exponent of the high-mass end of the clump/core 
mass function is found to be $\alpha_{\rm high} = -2.4 \pm 0.1$, 
consistent with the Salpeter mass function.  This result suggests that the 
shape of the stellar IMF may be apparent in the clump mass function on 
clump mass scales of several tens of $M_{\odot}$ or more and linear scales 
of several tenths of a parsec.  It also suggests that molecular cloud 
cores may be converted into stars (or small multiple systems) on an 
essentially one-to-one basis.  This scenario is consistent with theories 
in which massive stars form by core collapse and disk accretion.  It also 
relaxes the need to invoke either coalescence or competitive accretion to 
explain the formation of massive stars, though these processes probably do 
occur at some level.  This result deserves further investigation, 
especially to constrain any instrumental or post-processing effects which 
may affect it.

	The similarities among the shapes of the various clump/core mass 
functions are confirmed to be statistically significant using 
non-parametric pair-wise comparisons of the CMFs.  In 44 of 55 
comparisons, after normalizing away the differences in their median 
masses, the clump/core mass functions are found to have a high likelihood 
of representing random samplings from the same parent distribution.  That 
the observational CMFs agree well with the CMF of clumps extracted from a 
simulation of self-gravitating turbulent gas suggests that turbulent 
fragmentation is one possible sufficient explanation for the shape of the 
clump/core mass function.

\begin{acknowledgements}

M.~A.~R. has been supported by an Ontario Graduate Scholarship in Science 
and Technology.  Both M.~A.~R. and C.~D.~W. are supported by the Natural 
Sciences and Engineering Research Council of Canada (NSERC).  M.~A.~R. 
would like to thank E. Feigelson and F. Motte for helpful discussions 
during the preparation of this manuscript.

\end{acknowledgements}

\end{document}